\UseRawInputEncoding 

\documentclass[aps,prd,nofootinbib,nobibnotes,notitlepage,twocolumn]{revtex4-2}
\usepackage{graphicx}
\usepackage{amssymb}
\usepackage{amsmath} 
\usepackage{color}
\usepackage{float}
\usepackage{ulem}
\usepackage{accents}

\usepackage[utf8]{inputenc}
\usepackage[T1]{fontenc}

\usepackage{dsfont}
\usepackage{bbm} 

\usepackage{amsfonts}
\usepackage[colorlinks=true,
pdfstartview=FitV,linkcolor=blue,
citecolor=blue,urlcolor=blue,breaklinks=true]{hyperref}
\usepackage{array}
\usepackage{float}
\usepackage{placeins}
\usepackage[dvipsnames]{xcolor}
\usepackage{csquotes}
\usepackage[makeroom]{cancel}
\usepackage{units}
\usepackage{fixmath}
\usepackage{bigints}
\usepackage{amssymb}
\usepackage{systeme}
\newcolumntype{C}[1]{>{\centering\arraybackslash}m{#1}}
\renewcommand{\eqref}[1]{\mbox{Eq.~(\ref{#1})}}

\definecolor{ForestGreen}{rgb}{0.13,0.55,0.13}

\makeatletter
\renewcommand*\l@section{\@dottedtocline{1}{0em}{1.5em}}
\renewcommand*\l@subsection{\@dottedtocline{1}{1.5em}{1.5em}}
\renewcommand*\l@subsubsection{\@dottedtocline{1}{3em}{1.5em}}
\makeatother

\usepackage{capt-of}

\usepackage[caption=false]{subfig}

\begin{document}

	\title{Bi-isotropic effects on hybrid surface polaritons in bilayer configurations}

	\author{Ariel Nonato$^{a,b}$}
	\email{ariel.nonato@ufma.br}
	\author{Pedro D. S. Silva$^{a}$}
	\email{pedro.dss@ufma.br; pdiegoss.10@gmail.com}
		\affiliation{$^a$Coordena\c{c}\~ao do Curso de Ci\^encias Naturais - F\'isica, Universidade Federal do Maranh\~ao, Campus de Bacabal, Bacabal, Maranh\~ao, 65700-000, Brazil}
		
	\affiliation{$^b$Programa de P\'{o}s-gradua\c{c}\~{a}o em F\'{i}sica, Universidade Federal do Maranh\~{a}o, Campus Universit\'{a}rio do Bacanga, S\~{a}o Lu\'is (MA), 65080-805, Brazil}

\begin{abstract}

In this work, we investigate the bi-isotropic effects in the formation and tunability of hybrid surface polaritons in bilayer configurations. In order to do that, we consider a heterostructure constituted with layers formed by a TI medium endowed with bi-isotropic constitutive relations and an AFM medium. Using the transfex matrix formalism, we derive general expressions for the dispersion relations of surface polaritonic modes that explicitly include the dependence on the bi-isotropic coupling parameter, and analyze their coupling to bulk magnon-polaritons in the AFM layer. As an illustration of application, we consider a heterostructure formed with Bi$_{2}$Se$_{3}$ interfaced with antiferromagnetic (AFM) materials that support terahertz-frequency magnons, specifically Cr$_{2}$O$_{3}$ and FeF$_{2}$. In the strong bi-isotropic coupling regime, the surface Dirac plasmon-phonon-magnon polariton (DPPMP) dispersion undergoes a pronounced redshift, accompanied by the suppression of the characteristic anticrossing between the Dirac plasmon and the phonon. This effect, observed consistently across all AFM materials considered, suggests a weakening of the hybrid interaction, possibly due to saturation or detuning mechanisms induced by the increased $\alpha$ parameter. Furthermore, we demonstrate that increasing the Fermi energy of the topological insulator enhances the surface plasmon, phonon contribution, and induces a blueshift of the DPPP branches, effectively bringing them closer to resonance with the magnon mode and thereby increasing the hybridization strength. Intriguingly, this redshift partially compensates the blueshift induced by a higher Fermi level, restoring the system to a weak-coupling regime analogous to that observed at lower Fermi energies. Our findings reveal that both the Fermi level and the bi-isotropic response offer independent and complementary control parameters for tuning the strength of light-magnon coupling in TI/AFM heterostructures, with potential implications for reconfigurable THz spintronic and photonic devices.

\end{abstract}


\maketitle


\section{Introduction}

In general, the propagation of electromagnetic waves is governed by Maxwell's equations, supplemented by the constitutive relations that describe the electromagnetic response of the medium \cite{Jackson, Zangwill}. Consequently, media with specific electromagnetic properties can exhibit unusual features in wave propagation and optical phenomena, such as isotropic birefringence in exotic systems with magnetic currents \cite{PedroPRD2020}, and reflectance exceeding unity \cite{Nishida, AlexPRB2024}. Furthermore, an electromagnetic wave propagating through a medium can give rise to hybrid light-matter states or polaritons \cite{Albuquerque, Cottam}. The latter have attracted much attention over the years since their prediction in the 1950s \cite{Hopfield, Huang} and the experimental confirmation of phonon-polaritons \cite{Henry, Porto, Scott}.

When confined to interfaces between media with distinct properties, surface polaritons arise, governed by the electromagnetic response of the media. In this context, the interaction of incident light and collective excitations at the surface gives rise to surface electromagnetic waves propagating along the interface. These include, for instance, plasmon-polaritons at metal or semiconductor surfaces \cite{Heinz, Hoffman}, surface phonon-polaritons in polar dielectrics \cite{Cottam, Mills, Gubbin, Hellman}, and magnon-polaritons in magnetic materials \cite{Lehmeyer, Bauer, Macedo, Hao}.

Surface plasmon polaritons (SPPs) have been extensively investigated in several contexts, including surface microscopy \cite{Somekh, Demetriadou, Bozhevolnyi}, Weyl semimetals \cite{Peluso, Bugaiko}, Dirac materials \cite{Chihun}, tilted Weyl/Dirac systems \cite{Jalali}, topological insulators \cite{Junjie-Qi, Wang}, and interfaces with anisotropic media \cite{Darinskii, Golenitskii, Lahktakia}. In this scenario, surface Dirac plasmon polaritons (DPPs) are hybrid electromagnetic modes that arise from the coupling of surface-confined charge oscillations with the electromagnetic field at the interface of topological insulators (TIs) \cite{Nikolaos, Shu-Chen}. These excitations are characterized by propagation along the interface and evanescent decay perpendicular to it. DPPs have attracted considerable attention for potential applications in terahertz (THz) sensing, sub-diffraction imaging, photodetection, and data storage due to their confinement and tunability in the THz regime~\cite{ref1,ref2,ref3,ref4,ref5}.

In parallel, magnons---collective excitations associated with spin precession in magnetically ordered systems---have emerged as promising candidates for low-power information processing and spintronic applications at micro- and nanoscales. Their capability to transport information without charge movement makes them especially appealing for quantum technologies and energy-efficient devices~\cite{ref6,ref7,ref8,ref9,ref10,ref11,ref12,ref13,ref14}, being relevant when considering their potential coupling with plasmonic excitations in TIs materials.

When materials hosting Dirac plasmons, such as three-dimensional TIs (e.g., Bi$_2$Se$_3$, Bi$_2$Te$_3$, and Sb$_2$Te$_3$), are interfaced with magnetic systems---particularly antiferromagnetic (AFM) materials---the interplay between spin and charge collective modes can give rise to novel hybrid excitations, plasmon-magnon polaritons, with rich and tunable optical properties~\cite{ref15,ref16}. These modes are especially relevant when the constituent materials exhibit compatible excitation energies in the THz spectral window, allowing for coherent coupling.

Historically, the exploration of such hybrid polaritonic modes was limited by the mismatch in energy scales between magnons and plasmons in conventional systems. However, recent advances in material synthesis, interface engineering, and heterostructure fabrication have enabled experimental access to platforms where such coupling can be probed. Several AFM materials have been identified as suitable candidates for these studies, including NiO, MnF$_2$, FeF$_2$~\cite{ref17,ref18,ref19,ref20}, as well as Cr$_2$O$_3$, a magnetoelectric antiferromagnet with a Néel temperature near room temperature, and Mn$_2$Au, a metallic antiferromagnet with THz-frequency spin dynamics and promising ultrafast switching capabilities~\cite{ref22,ref23,ref24,ref25}. These developments have significantly expanded the landscape for investigating coherent interactions between Dirac plasmons and magnons, opening new avenues for engineered polaritonic devices that exploit spin-charge hybridization.

Previous studies have explored the coupling between Dirac plasmons in graphene and magnons in AFM materials, such as models describing the interaction between a graphene sheet and an AFM layer \cite{ref15}. Furthermore, an extension of this analysis incorporating damping mechanisms for both magnons and plasmons has been reported \cite{ref26}. Despite these advancements, those studies did not focus on quantifying the coupling strength between Dirac plasmons and magnons, nor did they address how such coupling depends on the intrinsic physical properties of the constituent material---a relevant aspect for guiding the search for systems exhibiting experimentally accessible strong coupling.

Furthermore, in bulk TIs, the coupling between Dirac plasmons and lattice vibrations (phonons) significantly modifies the dispersion of surface Dirac plasmon polaritons, giving rise to hybrid Dirac plasmon-phonon-polariton modes. These modes differ markedly from the polaritons observed in two-dimensional materials such as graphene~\cite{ref28}. In chalcogenide TIs with rhombohedral symmetry, such as Bi$_2$Se$_3$, two dominant infrared-active phonon modes arise when the incident electric field is oriented perpendicular to the $c$ axis: the $\alpha$ (E$_u^{(1)}$) and $\beta$ (E$_u^{(2)}$) modes~\cite{ref29}. The $\alpha$-phonon response produces a pronounced variation in the dielectric function of Bi$_2$Se$_3$ within the THz spectral range relevant to this work.

In the last decades, bi-isotropic constitutive relations have been extensively studied in various contexts~\cite{ref30,ref31,ref32,ref33,ref34}, including their relevance to topological phases of matter~\cite{ref35,ref36,ref37,ref38,ref39,ref40,ref41} and axion electrodynamics~\cite{ref42,ref43,ref44}. In such media, the coupling between electric and magnetic fields can be described by a magnetoelectric parameter, also known as a Tellegen in photonics or bi-isotropic coupling parameter, which modifies the wave propagation characteristics and optical signatures \cite{AlexPRB2024}. When applied to topological insulators, this additional degree of freedom enables new effects for the electromagnetic response and surface mode structure, especially in the terahertz regime.

Motivated by these investigations on plasmon-polariton modes and bi-isotropic signatures, in this work, we investigate the effects of a bi-isotropic coupling parameter in the description of Dirac plasmon-phonon-polariton (DPP) modes. Such a framework allows one to analyze the influences of magnetoelectric-like interactions on the electromagnetic surface modes, which potentially give rise to new hybridized states or modify existing resonant features. 

This paper is outlined as follows. In Section~\ref{section-theory}, we present the theoretical framework used to investigate the interaction between a bi-isotropic medium and an antiferromagnet (AFM). Sec.~\ref{section-maxwell} discusses briefly aspects related to the fundamental principles, starting from Maxwell’s equations and boundary conditions. In Sec.~\ref{section-solutions}, we present the general solutions for electromagnetic modes propagating in each bulk medium, explicitly considering the effects of the bi-isotropic parameter. In Section~\ref{section-surface-matrix}, the scattering matrix formalism is used to derive the general dispersion relation governing surface polariton modes in layered heterostructures that incorporates the dependence on the bi-isotropic parameter. In Section~\ref{section-applications}, we apply this formalism to a bilayer system composed of Bi$_2$Se$_3$ as the TI and representative AFM materials. Section~\ref{section-emergence} provides a general discussion of the conditions under which Dirac plasmon--phonon--magnon polaritons (DPPMPs) emerge in such systems. We then investigate how the resulting dispersion relations are affected by different combinations of TI and AFM materials, focusing primarily on Bi$_2$Se$_3$/Cr$_2$O$_3$ and Bi$_2$Se$_3$/FeF$_2$ heterostructures (Secs.~\ref{section-cr2o3}--\ref{section-fef2}). Emphasis is placed on identifying the material parameters that enable an experimentally accessible and tunable coupling between the TI and AFM layers. Finally, Section~\ref{section-final-remarks} summarizes the main findings and outlines potential directions for future research.

\section{\label{section-theory}Framework for bulk and surface modes}

In the theoretical description of light--matter interactions, different levels of approximation are commonly employed depending on the physical phenomena of interest and the complexity of the system under study~\cite{ref45,ref46}. In classical models, the interacting excitations, such as surface plasmons, phonons, or magnons, are treated as coupled harmonic oscillators, where the coupling strength is introduced as an adjustable parameter. This approach is useful for fitting experimental dispersion relations, but it does not explicitly reveal how the coupling arises from microscopic properties or material parameters \cite{ref47}.

In a semiclassical approach, one combines Maxwell's equations with the frequency-dependent optical responses of each medium. This formalism allows for a direct connection between hybridization of modes the the intrinsic electric and magnetic properties of the constituent materials. The resulting eigenmodes, known as polaritons, arise naturally from the boundary conditions and continuity requirements imposed at interfaces between distinct materials. Within this framework, the coupling strength emerges as a consequence of field matching and material contrast, without the need for empirical parameters. Finally, in the quantum mechanical description, the hybrid modes arise as coherent superpositions of light and matter excitations, with the coupling governed by interaction terms in the Hamiltonian. This framework enables the study of quantum coherence, strong coupling, and entanglement. Nevertheless, when analyzing macroscopic dispersion and identifying modes in layered media, the semiclassical approach offers a reliable framework for effective description.

In this work, we adopt the semiclassical formalism to study the formation and dispersion of surface Dirac plasmon--phonon--magnon polaritons (DPPMPs) in a heterostructure composed bi-isotropic medium and an antiferromagnetic (AFM) material. We solve Maxwell’s equations for this layered system using transfer matrix and scattering matrix techniques, which allow us to investigate wave propagation and field continuity across multiple interfaces. The hybrid modes are obtained by imposing boundary conditions, yielding the corresponding evanescent surface waves. Here, we consider a bi-isotropic medium, in which the bi-isotropic parameter characterizes the Tellegen-type response of the TI and leads to a coupling between the electric and magnetic fields \cite{ref48}. Rather than introducing anisotropy or tensorial magnetoelectric effects, it acts as an isotropic pseudoscalar term that modifies the electromagnetic response of the medium. This modification enables a systematic investigation of how bi-isotropic behavior, distinct from conventional magnetoelectric coupling, affects the hybridization strength and spectral features of the DPPMPs. In particular, it allows one to examine how bi-isotropic effects influence the dispersion of the DPPMP modes, particularly the strength of the hybridization near the magnon resonance.

Prior to presenting the numerical results, we describe our computational framework: we consider a heterostructure composed of $N$ constituent layers and utilize state-of-the-art transfer and scattering matrix methods that are both robust and scalable. These methods accommodate arbitrary complexity in material stacking and dielectric/magnetic contrast, enabling accurate determination of the surface polariton resonance conditions~\cite{ref49,ref51,ref53,ref54}.

Unlike the model described in Ref.~\cite{ref15}, which applies the transfer matrix technique only to the specific case of graphene coupled to an AFM layer, or the formulation in Ref.~\cite{ref56}, which describes surface plasmon modes in multilayers but faces numerical instability and convergence issues when treating many finite-thickness layers, our approach, following Ref.~\cite{ref55}, provides a more general formulation. Within this framework, Maxwell’s equations can be solved to determine the dispersion relations of the surface DPPMPs in complex layered heterostructures. Thus, one can find a consistent description of the surface polaritons—evanescent electromagnetic waves that decay along the propagation direction—which leads to the general dispersion equation governing the surface polariton modes in the heterostructure.

\subsection{\label{section-maxwell}Description of the system and constitutive relations}

\begin{figure}[h]
\begin{centering}
\includegraphics[scale=0.47]{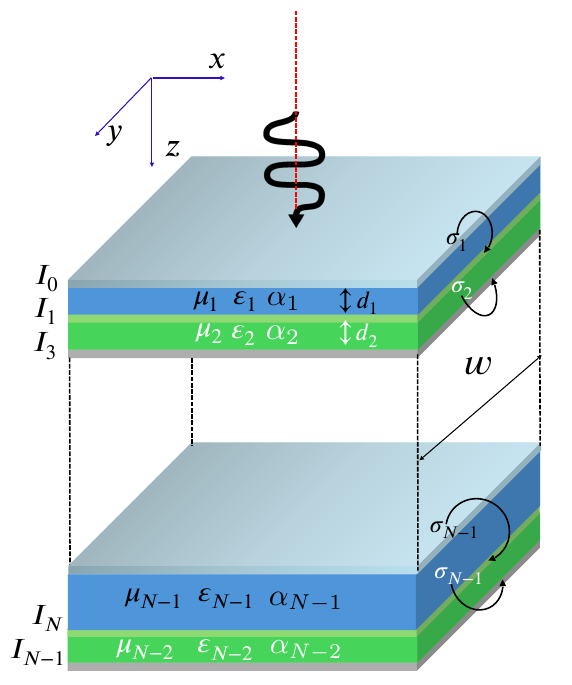}
\par\end{centering}
\caption{\small{\label{figure-bilayers-with-alpha-2}}Schematic of a multilayer structure consisting of $N$ constituent layers that have the same width $w$ along the $x$-direction. The $z$-axis is chosen as the growth direction of the structure. The thickness, premittivity, permeability, bi-isotropic parameter in the $m$th layer, and optical conductivity of the carrier sheet at the $m$th surface/interface are denoted by $d_{m}$, $\epsilon_{m}$, $\mu_{m}$, $\alpha_{m}$ and $\sigma_{m}$, respectively, whereas $I_{m}$ indicates the interface matrix at the $m$th interface.}
\end{figure}

\begin{figure}[h]
\begin{centering}
\includegraphics[scale=0.46]{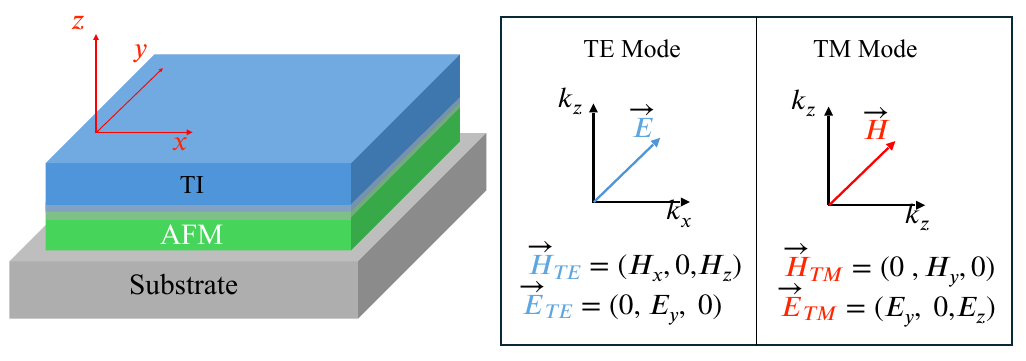}
\par\end{centering}
\caption{\small{\label{figure-structure-bilayer}}Schematic of a multilayer structure consisting of $N$ layers. Electric and magnetic field components. The $z$-axis is chosen as the growth direction of the structure. The propagation of the electromagnetic wave is confined in $x$-$z$ plane.}
\end{figure}

We analyze a multilayer heterostructure composed of distinct layers, with an incident electromagnetic wave impinging from the top side of the structure (see Fig.~\ref{figure-bilayers-with-alpha-2}). The growth direction is defined along the $z$-axis. The system is considered to be infinitely extended along the $y$-axis, while it has a finite thickness in the $x$-direction, comprising successive layers of a topological insulator (TI), an antiferromagnetic material (AFM), and a substrate. As shown in Fig.~\ref{figure-structure-bilayer}, the propagation of the electromagnetic wave is confined to the $x$--$z$ plane and directed along the positive $z$ direction. In this case, for $s$-polarized incident wave, the electric field vector is oriented along the $y$-axis, whereas for $p$-polarized wave, the magnetic field component lies along the same axis.

\begin{figure}[h]
\begin{centering}
\includegraphics[scale=0.5]{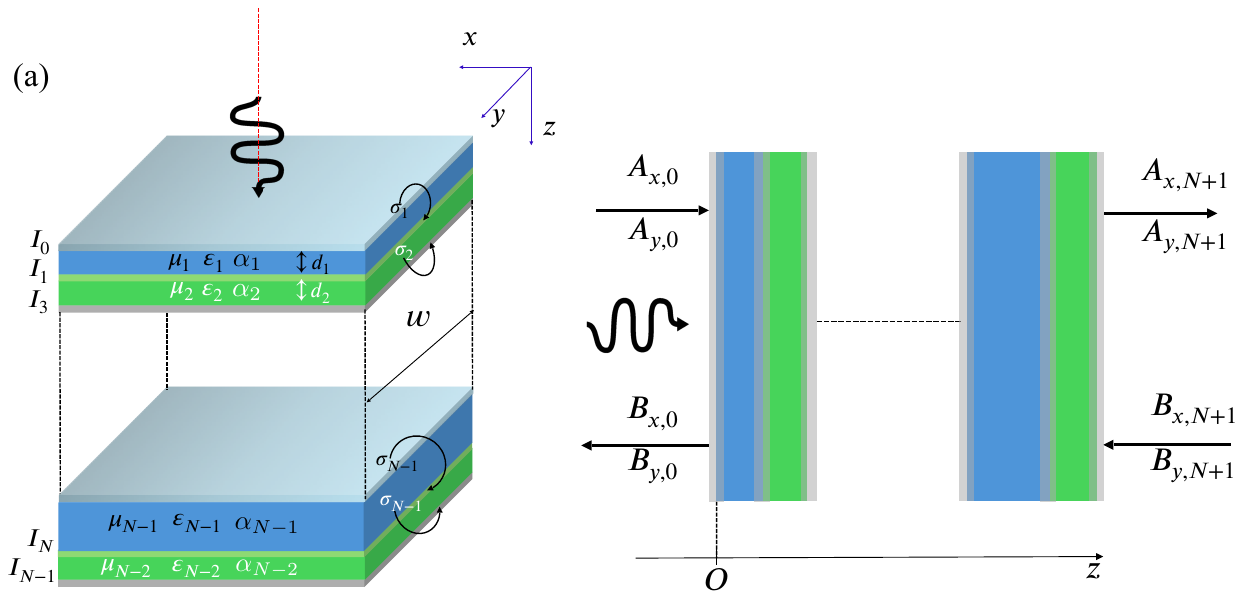}
\par\end{centering}
\caption{\small{\label{figure-incoming-outgoing-fields}} Schematic of the amplitudes of incoming and outgoing EM waves used in the scattering matrix approach. The EM wave is incident on the left of the surface in the figure.}
\end{figure}

The electromagnetic wave propagating through each region of the heterostructure satisfies Maxwell's equations, with standard electromagnetic boundary conditions applied at the interfaces between adjacent media. Assuming the absence of free charges and currents in the bulk, Maxwell’s equations take the form:
\begin{align}
\nabla \cdot \mathbf{D} &= 0, \quad   \nabla \times \mathbf{H} = \frac{\partial \mathbf{D}}{\partial t} , \label{eq-maxwell-1} \\
 \nabla \cdot \mathbf{B} &= 0, \quad \hspace{0.1cm} \nabla \times \mathbf{E} = -\frac{\partial \mathbf{B}}{\partial t} .  \label{eq-maxwell-2}
\end{align}

The system is governed by modified constitutive relations that account for the response characteristics of bi-isotropic linear media. These relations can be written as
\begin{align}
\mathbf{D} &= \epsilon_{0} \epsilon \mathbf{E} + \mu_{0} \alpha \mathbf{H}, \label{eq-constitutive-relation-1} \\
\mathbf{B} &= \mu_{0} \mu \mathbf{H} + \mu_{0} \alpha \mathbf{E}, \label{eq-constitutive-relation-2}
\end{align}
where $\mathbf{E}$, $\mathbf{D}$, $\mathbf{B}$, and $\mathbf{H}$ denote the electric field, electric displacement field, magnetic induction, and magnetic field, respectively. The constants $\varepsilon_0$ and $\mu_0$ are the permittivity and permeability of free space, and $\epsilon$ and $\mu$ represent the relative permittivity and permeability of the material. The parameter $\alpha$ is the bi-isotropic coefficient, which captures the linear cross-response permitted in such media due to their symmetric and isotropic electromagnetic behavior. The boundary conditions at the $m$th interface are given by:

\begin{align}
\mathbf{n} \times \left( \mathbf{H}_{m+1} - \mathbf{H}_m \right) \big|_{z = z_m} &= \mathbf{J}_m,  \label{eq-boundary-1} \\
\mathbf{n} \times \left( \mathbf{E}_{m+1} - \mathbf{E}_m \right) \big|_{z = z_m} &= 0, \label{eq-boundary-2}\\
\mathbf{n} \cdot \left( \mathbf{D}_{m+1} - \mathbf{D}_m \right) \big|_{z = z_m} &= \rho_m, \label{eq-boundary-3}\\
\mathbf{n} \cdot \left( \mathbf{B}_{m+1} - \mathbf{B}_m \right) \big|_{z = z_m} &= 0,  \label{eq-boundary-4}
\end{align}
where $\mathbf{n}$ is a unit vector perpendicular to the $m$th interface, $\mathbf{J}_m$ is the in-plane current, and $\rho_m$ is the carrier density of the electron gas at the $m$th interface.

Taking the curl of second expressions in Eqs.~(\ref{eq-maxwell-1}) and (\ref{eq-maxwell-2}), and implementing the constitutive relations given in \eqref{eq-constitutive-relation-1} and \eqref{eq-constitutive-relation-2} into first expressions of \eqref{eq-maxwell-1} and \eqref{eq-maxwell-2}, one obtains the following set of coupled wave equations for the electromagnetic fields
\begin{align}
\left[\left(\frac{\mu\epsilon}{c^{2}} - \mu_{0}^{2}\alpha^{2} \right) \partial_{t}^{2} + \partial_{i}\partial_{j} - \partial_{j}^{2} \right] F^{j} &=0,  \label{eq-wave-1} 
\end{align}
where $F^{j}=(E^{j}, H^{j})$, and
\begin{align}
\nabla \cdot (\mu_0 \alpha \mathbf{E}) &= -\mu_0 \mu \, \nabla \cdot \mathbf{H},  \label{eq-wave-2}\\
\nabla \cdot (\mu_0 \alpha \mathbf{H}) &= -\varepsilon_0 \varepsilon \, \nabla \cdot \mathbf{E}. \label{eq-wave-3}
\end{align}

These wave equations form the basis for determining the dispersion relation, which connects the energy (or frequency) of the electromagnetic wave in the material to its wavevector. 

Considering the propagation of a plane electromagnetic wave (e.g., using an ansatz of the form ${\bf{E}}$, ${\bf{H}}$ $\propto e^{i({\bf{k}}\cdot {\bf{r}} - \omega t)}$), in \eqref{eq-wave-1}, one obtains
\begin{align}
\left[ \partial_{z}^{2} - k_{x,m}^{2} + \left( \frac{\mu \varepsilon}{c^2} - \mu_{0}^{2} \alpha^{2} \right) \omega^{2} \right] \mathbf{E} = {\bf{0}}. \label{eq-wave-5}
\end{align}
The solution must satisfy:
\begin{align}
\left[\partial_{z}^{2} + k_{z,m}^{2} \right] \mathbf{E} ={\bf{0}}, \label{eq-wave-6}
\end{align}
where
\begin{align}
k_{z,m}^{2} = \left[
\left( \frac{\mu \varepsilon}{c^{2}} - \mu_{0}^{2} \alpha^{2}_{m} \right) \omega^{2} - k_{x,m}^{2} \right] , \label{eq-wave-7}
\end{align}

The expression for $k_{z,m}$ presented in \eqref{eq-wave-7} corresponds to the modified longitudinal component of the wavevector in the $m$th layer. The presence of $\mu_{0}^{2}\alpha^{2}$ introduces a correction to the standard dispersion relation due to cross-coupling between the electric and magnetic fields, as typically observed in media exhibiting axion-like or Tellegen-type responses.

In the following section, we present detailed analytical solutions of these equations in order to derive the bulk polariton modes supported by each medium, namely the topological insulator (TI) with a bi-isotropic parameter and the antiferromagnet (AFM).

\subsection{\label{section-solutions}Dispersion relations for bulk propagating modes}

We now focus on the bulk polariton modes supported within each constituent material of the heterostructure depicted in Fig.~\ref{figure-bilayers-with-alpha-2} and Fig.~\ref{figure-incoming-outgoing-fields}, which illustrates the general incoming and outgoing fields in the system. Since the electromagnetic wave propagates in the $x$-$z$ plane, the solutions  to equations (\ref{eq-wave-1}) -- (\ref{eq-wave-3}) within the $m$th bulk layer, as illustrated in Fig.~\ref{figure-structure-bilayer}, can be explicitly expressed as
\begin{align}
\mathbf{E}_{m} &= e^{i(k_{x,m} x - \omega t)}
\mathcal{F}
\begin{pmatrix}
A_{x,m} \\
A_{y,m} \\
B_{x,m} \\
B_{y,m}
\end{pmatrix} , \label{eq-field-1} \\
{\bf{H}}_{m} &= e^{i(k_{x,m}x-\omega t)} \mu_{0} \zeta_{m} \left(\frac{}{} \mathcal{G} - \mu_{0} \alpha_{m} \mathcal{F} \right)  \begin{pmatrix}
A_{x,m} \\
A_{y,m} \\
B_{x,m} \\
B_{y,m}
\end{pmatrix} , \label{eq-field-1-1}
\end{align}
with
\begin{align}
\mathcal{F} &=\begin{bmatrix}
e^{i k_{z,m} z} & 0  & e^{-i k_{z,m} z} & 0 \\
0 & e^{i k_{z,m} z} & 0 & e^{-i k_{z,m} z} \\
-\frac{k_x}{k_{z,m}} e^{i k_{z,m} z} & 0 & \frac{k_x}{k_{z,m}} e^{-i k_{z,m} z} & 0
\end{bmatrix} , \label{eq-field-2}
\end{align}
\begin{align}
\mathcal{G}&= e^{-i k_{z,l}z}\begin{bmatrix}
0 & - \frac{k_{z,l}}{\omega} e^{2i k_{z,l}z}& 0 & \frac{k_{z,l}}{\omega}  \\
\frac{\kappa_{l}}{\omega}  e^{2i k_{z,l}z}& 0 & - \frac{\kappa_{l}}{\omega} & 0 \\
0 & k_{x,l} e^{2ik_{z,l}z} &0 &  k_{x,l} 
\end{bmatrix} , \label{eq-field-3}
\end{align}
\begin{align}
\kappa_{m} &= \frac{k_{x,m}^{2}+ k_{z,m}^{2}}{k_{z,m}} , \label{eq-field-4}
\end{align}
where $A_{j,m}$ and $B_{j,m}$ represent the amplitudes of the $j$ ($j=x, y$)  components of the forward and backward propagating electromagnetic wave in the $m$th layer. The spatial variables $x$ and $z$ refer to the Cartesian coordinates in the corresponding directions.

Substituting \eqref{eq-field-1} into \eqref{eq-wave-5},  one obtains
\begin{align}
M_{ij} X^{j} &=0 , \label{eq-field-5}
\end{align}
where
\begin{align}
M_{ij}&=\left(k_{x,m}^{2} + k_{z,m}^{2} -  \epsilon_{m}\mu_{m} \frac{\omega^{2}}{c^{2}} + \mu_{0}^{2}  \alpha_{m}^{2} \right) \delta_{ij}, \label{eq-field-5-1}\\
X^{i} &= \mathcal{F}_{ij} Y^{j}, \quad {\bf{Y}} =  (A_{x,m}, A_{y,m}, B_{x,m}, B_{y,m} )^{T} . \label{eq-field-6}
\end{align}
Here, $\epsilon_{m}$, $\mu_{m}$, and $\alpha_{m}$ are the electric permittivity, magnetic permeability, and the bi-isotropic parameter associated with the $m$th medium in the structure. The non-trivial solutions of \eqref{eq-field-5} are obtained by requiring $\mathrm{det}[M_{ij}]=0$. The latter condition allows us to determine the dispersion relations for the propagating modes in the system.


\subsection{\label{section-surface-matrix}Surface polaritons modes}

To analyze the surface polaritons, we implement the boundary conditions at the $m$th interface of the structure in Fig.~\ref{figure-bilayers-with-alpha-2}, which allows us to determine the dispersion of surface modes, which are characterized by confinement to the interface and evanescent decay perpendicular to the interface. 

Considering the normal to the interface as ${\bf{n}}=(0, 0, 1)$, and the conditions
\begin{subequations}
\label{eq-condition-0}
\begin{align}
\mathbf{n} \times \left( \mathbf{H}_{m+1} - \mathbf{H}_m \right)\big|_{z = z_m} &= \mathbf{J}_m, \label{eq-condition-1}\\
\mathbf{n} \times \left( \mathbf{E}_{m+1} - \mathbf{E}_m \right)\big|_{z = z_m} &= 0, \label{eq-condition-2}
\end{align}
\end{subequations}
\begin{align}
 \mathbf{J}_m &= \boldsymbol{\sigma}_m \mathbf{E}_{m+1}, &
\quad \boldsymbol{\sigma}_m &= 
\begin{pmatrix}
\sigma_m^{xx} & \sigma_m^{xy} \\
\sigma_m^{yx} & \sigma_m^{zz}
\end{pmatrix}, \label{eq-condition-3}
\end{align}
where $\mathbold{\sigma}_{m}$ is the optical conductivity tensor of the corresponding two-dimensional carrier gas at the $m$th interface. Implementing Eqs.(\ref{eq-field-1}) and (\ref{eq-field-1-1}) into \eqref{eq-condition-0}, one finds (after some algebra)
\begin{align}
\begin{pmatrix}
A_{x,m} \\
A_{y,m} \\
B_{x,m} \\
B_{y,m}
\end{pmatrix}
= \mathcal{I}_m
\begin{pmatrix}
A_{x,m+1} \\
A_{y,m+1} \\
B_{x,m+1} \\
B_{y,m+1}
\end{pmatrix} , \label{eq-interface-matrix-1}
\end{align}
where $\mathcal{I}_{m}$ is the interface matrix connecting the amplitudes of the electromagnetic waves in the adjacent $m$th and $(m+1)$th layers. The matrix $\mathcal{I}_{m}$ is given by
\begin{align}
\mathcal{I}_m = 
\begin{pmatrix}
\mathcal{C} \\
\mathcal{U}_{m}
\end{pmatrix}^{-1}
\begin{pmatrix}
\mathcal{C} \\
\mathcal{V}_{m}
\end{pmatrix} , \label{eq-interface-matrix-2}
\end{align}
where
\begin{align}
\mathcal{U}_{m} &= -\dfrac{\mathcal{Q}}{\mu_0} \zeta_{m} 
\begin{bmatrix}
\alpha_{m} \mu_0 & \dfrac{k_{z,m}}{\omega} & \alpha_m \mu_0 & -\dfrac{k_{z,m}}{\omega} \\
-\dfrac{\kappa_{m}}{\omega } & \alpha_{m} \mu_{0} & \dfrac{\kappa_{m}}{\omega } & \alpha_{m} \mu_{0} \\
0 & -\dfrac{k_{x,m}}{\omega} & 0 & \dfrac{k_{x,m}}{\omega}
\end{bmatrix} ,\label{eq-interface-matrix-3}\\
\mathcal{V}_{m} &= \frac{\mathcal{Q} }{\mu_0}\zeta_{m+1} \mathcal{J}_{m} + \mathcal{L}_{m}, \label{eq-interface-matrix-4} 
\end{align}
with $\zeta_{m}=\mu^{-1}_{m}$, 
\begin{align}
\mathcal{J}_{m} &= \begin{bmatrix}
\alpha_{m+1} \mu_{0} & -\frac{k_{z,m+1}}{\omega} & \alpha_{m+1} \mu_{0} & \frac{k_{z,m+1}}{\omega} \\
\frac{\kappa^{\prime}_{m}}{\omega} & - \alpha_{m+1}\mu_{0} & - \frac{\kappa^{\prime}_{m}}{\omega} & -\alpha_{m+1} \mu_{0} \\
0 & \frac{k_{x,m+1}}{\omega} & 0 & - \frac{k_{x,m+1}}{\omega} 
\end{bmatrix} , \label{eq-interface-matrix-5} \\
\mathcal{L}_{m} &= \begin{bmatrix}
- \sigma^{yx}_{m} & - \sigma^{yy}_{m} & -\sigma^{yx}_{m} & - \sigma^{yy}_{m} \\
\sigma^{xx}_{m} & \sigma^{xy}_{m} & \sigma^{xx}_{m} & \sigma^{xy}_{m} 
\end{bmatrix} , \label{eq-interface-matrix-6} \\
\mathcal{C} &= \begin{bmatrix}
1 & 0 & 1 & 0 \\
0 & 1 & 0 &1
\end{bmatrix} , \quad \mathcal{Q} = \begin{bmatrix}
1 & 0 & 0 \\
0 & 1 & 0
\end{bmatrix} , \label{eq-interface-matrix-7} \\
\kappa^{\prime}_{m} &= \frac{k_{z, m+1}^{2} + k_{x,m+1}^{2}}{ k_{z,m+1}} , \quad \kappa_{m} = \frac{k_{z, m}^{2} + k_{x,m}^{2}}{ k_{z,m}}  .  \label{eq-interface-matrix-8}
\end{align}

And the propagation matrix between the $m$th and $(m+1)$th interfaces can be defined by
\begin{align}
\mathcal{P}_{m,m+1} &= \begin{pmatrix}
e^{-ik_{z,m+1}d_{m+1}}  \mathbbm{1}_{2}& 0 \\
0 & e^{i k_{z,m+1}d_{m+1}}  \mathbbm{1}_{2} 
\end{pmatrix} ,
\end{align}
with $d_{m+1}$ being the thickness of $(m+1)$th layer.

\section{\label{section-applications}Application to TI/AFM bilayers with bi-isotropic parameter}

We now apply the theoretical framework developed in the previous section to investigate the electromagnetic interactions between a topological insulator (TI) and an antiferromagnetic (AFM) material. In our model, the essential input parameters include the thicknesses of each constituent layer, the frequency-dependent dielectric permittivities, the corresponding magnetic permeability tensors, and the bi-isotropic parameter. Additionally, we incorporate the optical conductivities associated with the two-dimensional electronic carriers, both on the surface of the topological insulator and at the TI/AFM interface. These quantities play a fundamental role in capturing the key physical mechanisms that govern the hybridization and dispersion of the polariton modes at the interface, especially in systems where topological surface states and magnetic ordering coexist.

Firstly, in the absence of an external magnetic field, the optical conductivity tensor of the two-dimensional carrier gas, both at the surface and at the interface between the two materials, assumes a diagonal form
\begin{align}
\sigma_{ij} &= \sigma \delta_{ij}. \label{eq-application-1}
\end{align}

The magnetic permeability tensor of uniaxial or cubic antiferromagnetic layers considered in this work can be written as~\cite{ref57}
\begin{align}
\mu_{ij} &= \mu_{\xi \xi} \delta_{ij} . \label{eq-application-2}
\end{align}
The quantity $\mu_{\xi\xi}$, with $\xi \in \{x, y, z\}$, depends on the orientation of the static magnetization vector within the antiferromagnetic (AFM) medium. Specifically, when the magnetization is aligned along the $\xi$, the corresponding component of the permeability tensor reduces to the vacuum value, $\mu_{\xi\xi} = 1$, reflecting the absence of a dynamic magnetic response in that direction. Conversely, for directions orthogonal to the magnetization axis, the magnetic permeability exhibits a resonant behavior characteristic of antiferromagnetic order, and is given by
\begin{align}
\mu_{\xi\xi} = 1 + 4\pi \frac{2 \gamma^2 H_a M}{\Omega_0^2 - \left( \omega + \dfrac{i}{\tau_{\text{mag}}} \right)^2} , \label{eq-application-3} 
\end{align}
where $H_{a}$ is the magnetic anisotropy field, $M$ denotes the sublattice magnetization, and $\gamma = \frac{g e}{2mc}$ is the gyromagnetic ratio\footnote{Expressed in CGS units.}, with $g$ being the Landé $g$-factor, $e$ the elementary charge, $m$ the electron mass, and $c$ the speed of light.

The parameter $\Omega_0$ defines the antiferromagnetic resonance (AFMR) frequency, given by:
\begin{align}
\Omega_0 = \gamma \sqrt{2(H_a + H_e) H_a} , \label{eq-application-4}
\end{align}
where $H_e$ is the exchange field responsible for coupling between magnetic sublattices.
The denominator in the expression for $\mu_{\xi\xi}$ includes a complex frequency term $\omega + i/(\tau_{\text{mag}})$, which accounts for dissipative damping in the magnetic response, with $\tau_{\text{mag}}$ representing the magnetic relaxation time. This formulation captures both the resonant enhancement of the permeability near the AFMR frequency and the losses due to damping mechanisms inherent in real antiferromagnetic systems.
The dielectric function of corresponding layers, which are isotropic materials considered in this work, in the structure is given by the Drude–Lorentz model~\cite{ref58}:

\begin{align}
\varepsilon^{TI} = \varepsilon_{\infty} + \frac{\omega_p^2}{\omega^2 + i\omega \Gamma}
+ \sum_{n=1}^{N} \frac{\omega_{p,n}^2}{\omega_{0,n}^2 + \omega^2 + i\omega \Gamma_n} , \label{eq-application-5}
\end{align}
where $\varepsilon_{\infty}$ is the dielectric constant at high frequency ($\omega \to \infty$), the second term on the right-hand side of \eqref{eq-application-5} indicates the Drude bulk contributions, and the third term is a sum of all contributions from the other Lorentz oscillators present in the system.

\subsection{\label{section-emergence}Formation of surface polariton mode in the bi-isotropic structure}

\begin{figure}
  \centering
  \subfloat{\includegraphics[width=0.22\textwidth]{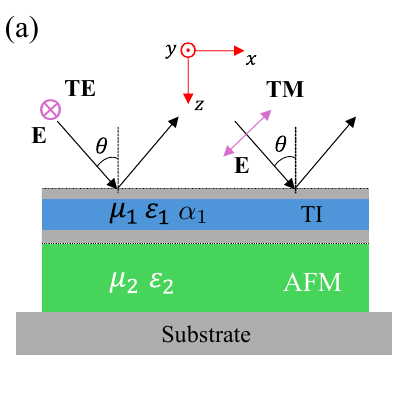}%
    \label{figure-bilayer-fields-amplitudes-A}}
  \hfill
  \subfloat{\includegraphics[width=0.22\textwidth]{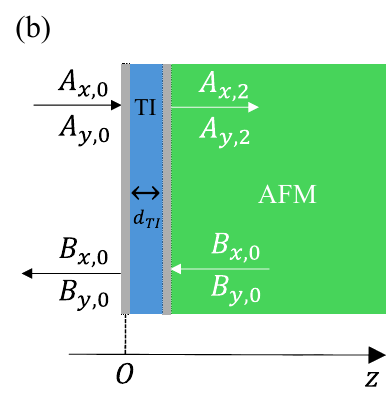}%
    \label{figure-bilayer-fields-amplitudes-B}}
  \caption{\small{\label{figure-bilayer-fields-amplitudes}}(a) The structure consisting of a topological insulator (TI) and an antiferromagnet (AFM) with an arbitrary magnetization direction. An electromagnetic (EM) wave, generally containing both TE and TM polarizations, impinges on the surface of the TI material to induce electric excitations in the TI film, which can couple to magnetic modes in the AFM layer. (b) A finite TI film of thickness $d_{\mathrm{TI}}$ is placed on a semi-infinite AFM substrate. 
Here, $A_{i,j}$ and $B_{i,j}$, with $i = x, y$ and $j = 0, 2$, denote the amplitudes 
of the forward- and backward-propagating electromagnetic waves in air ($j = 0$) and in the AFM medium ($j = 2$).
}
\end{figure}

We first discuss the conditions that enable the formation of surface Dirac plasmon–phonon–magnon polaritons (DPPMPs) in the TI/AFM heterostructure illustrated in Fig.~\ref{figure-bilayer-fields-amplitudes}. The system consists of a topological insulator (TI) thin film deposited on an antiferromagnetic (AFM) layer, which is supported by an MgO substrate. When an electromagnetic (EM) wave with mixed TM and TE polarization impinges on the TI surface, it can excite collective charge and spin dynamics: Dirac plasmons confined to the TI interfaces and magnons associated with spin-wave modes in the AFM. The electrodynamic coupling between these modes gives rise to hybrid quasiparticles—DPPMPs— which are characterized by modifications in the dispersion relation $\omega(k)$. Additionally, the effects of the bi-isotropic parameter are encoded into the general dispersion relation of the surface polaritons. In the following, we examine the effects that the bi-isotropic parameter can yield in the spectral behavior of these hybrid excitations.

For simplicity, the antiferromagnetic (AFM) layer is assumed to be thick enough to be treated as semi-infinite. As shown in  Fig.~\ref{figure-bilayer-fields-amplitudes}\subref{figure-bilayer-fields-amplitudes-B}, the configuration consists of a TI film of finite thickness $d_{\mathrm{TI}}$ in contact with a semi-infinite AFM. The symbols $A_{i,j}$ and $B_{i,j}$, where $i = x, y$ and $j = 0, 2$, represent the amplitudes of EM waves moving forward and backward in the air ($j = 0$) and in the AFM layer ($j = 2$).

To analyze the dispersion of surface Dirac plasmon–phonon–magnon polaritons (DPPMPs) in this system, we
consider the case in which the amplitudes of the electromagnetic fields satisfy the conditions $A_{x,0} = A_{y,0} = B_{x,2} = B_{y,2} = 0$. Under these conditions, the total transfer matrix $\mathcal{T}$, which governs the continuity of the fields across the
structure, can be written as $\mathcal{T} = \mathcal{I}_{0} \mathcal{P} \mathcal{I}_{1}$, where $\mathcal{I}_{0}$ is the interface matrix between air and the TI, $\mathcal{I}_{1}$ is the interface matrix
between the TI and the AFM, and $\mathcal{P}$ is the propagation matrix within the TI layer. The transfer-matrix ($\mathcal{T}$) method is widely recognized for its robustness and efficiency in modeling the optical response of multilayered structures~\cite{ref59}. The transfer matrix formalism adopted here follows procedures discussed in previous works~\cite{ref55,ref49}. To construct this matrix $\mathcal{T}$, using \eqref{eq-application-2} in Eqs.~(\ref{eq-interface-matrix-3}) and (\ref{eq-interface-matrix-4}), one finds the interface matrix $\mathcal{I}_{m}$ as
\begin{align}
\mathcal{I}_{m} &= \begin{bmatrix}
\mathcal{K}_{m} && \mathcal{M}_{m} \\
\\
\mathcal{N}_{m} && \mathcal{R}_{m}
\end{bmatrix} , \label{eq-interface-matrix-general-1}
\end{align}

\begin{widetext}
where
\begin{subequations}
\label{eq-interface-matrix-general-2}
\begin{align}
\mathcal{K}_{m} &= \begin{bmatrix}
1+ \dfrac{\mu_{m}^{y}}{\mu_{m+1}^{y}} \dfrac{\kappa^{\prime}_{m}}{\kappa_{m}} + \dfrac{\omega \mu_{0} \mu_{m}^{y} \sigma_{m}}{\kappa_{m}} && \dfrac{\omega \mu_{0}}{\kappa_{m} \mu_{m+1}^{y}} (\mu_{m+1}^{y} \alpha_{m} - \mu_{m}^{y} \alpha_{m+1}) \\
\\
\dfrac{-\omega \mu_{0}}{k_{z,m} \mu_{m+1}^{x}} (\mu_{m+1}^{x} \alpha_{m} - \mu_{m}^{x} \alpha_{m+1}) && 1+ \dfrac{\mu_{m}^{x} k_{z,m+1}}{\mu_{m+1}^{x} k_{z,m}} + \dfrac{\omega \mu_{0} \mu_{m}^{x} \sigma_{m}}{k_{z,m}}
\end{bmatrix} ,  \label{eq-interface-matrix-general-3} 
\end{align}

\begin{align}
\mathcal{M}_{m} &=  \begin{bmatrix}
1- \dfrac{\mu_{m}^{y}}{\mu_{m+1}^{y}} \dfrac{\kappa^{\prime}_{m}}{\kappa_{m}} + \dfrac{\omega \mu_{0} \mu_{m}^{y} \sigma_{m}}{\kappa_{m}}  && \dfrac{\mu_{0}\omega}{\kappa_{m} \mu_{m+1}^{y}} (\mu_{m+1}^{y} \alpha_{m} - \mu_{m}^{y} \alpha_{m+1}) \\
\\ 
\dfrac{-\omega \mu_{0}}{k_{z,m} \mu_{m+1}^{x}} (\mu_{m+1}^{x} \alpha_{m} - \mu_{m}^{x} \alpha_{m+1})  && 1- \dfrac{\mu_{m}^{x} k_{z,m+1}}{\mu_{m+1}^{x} k_{z,m}} + \dfrac{ \omega \mu_{0}\mu_{m}^{x} \sigma_{m}}{k_{z,m}}
\end{bmatrix} , \label{eq-interface-matrix-general-4} 
\end{align}

\begin{align}
\mathcal{N}_{m} &= \begin{bmatrix}
1- \dfrac{\mu_{m}^{y} }{\mu_{m+1}^{y}} \dfrac{\kappa^{\prime}_{m}}{\kappa_{m}} - \dfrac{\omega \mu_{0} \mu_{m}^{y} \sigma_{m}}{\kappa_{m}} &&  \dfrac{\omega \mu_{0}}{\kappa_{m} \mu_{m+1}^{y}} (\mu_{m}^{y} \alpha_{m+1}-\mu_{m+1}^{y} \alpha_{m})  \\
\\
\dfrac{-\omega \mu_{0}}{k_{z,m} \mu_{m+1}^{x}} (\mu_{m}^{x} \alpha_{m+1}- \mu_{m+1}^{x} \alpha_{m} ) &&  1- \dfrac{ \mu_{m}^{x} k_{z,m+1}}{\mu_{m+1}^{x} k_{z,m}} - \dfrac{\omega \mu_{0} \mu_{m}^{x} \sigma_{m}}{k_{z,m}} 
\end{bmatrix}  , \label{eq-interface-matrix-general-5} 
\end{align}

\begin{align}
\mathcal{R}_{m} &= \begin{bmatrix}
1+\dfrac{\mu_{m}^{y}}{\mu_{m+1}^{y}} \dfrac{\kappa^{\prime}_{m}}{\kappa_{m}} - \dfrac{\omega \mu_{0} \mu_{m}^{y} \sigma_{m}}{\kappa_{m}} && \dfrac{\omega \mu_{0}}{\kappa_{m}\mu_{m+1}^{y}} (\mu_{m}^{y}\alpha_{m+1} - \mu_{m+1}^{y} \alpha_{m} ) \\
\\ 
\dfrac{-\omega \mu_{0}}{k_{z,m} \mu_{m+1}^{x}} (\mu_{m}^{x} \alpha_{m+1} - \mu_{m+1}^{x} \alpha_{m}) && 1 + \dfrac{ \mu_{m}^{x} k_{z,m+1}}{\mu_{m+1}^{x} k_{z,m}}  - \dfrac{\omega \mu_{0} \mu_{m}^{x} \sigma_{m}}{k_{z,m}} 
\end{bmatrix} , \label{eq-interface-matrix-general-6} 
\end{align}
\end{subequations}
\end{widetext}
with $\kappa_{m}^{\prime}$ and $\kappa_{m}$ are given in \eqref{eq-interface-matrix-8}.

Using the explicit forms of the interface matrices $\mathcal{I}_{0}$ and $\mathcal{I}_{1}$, together with the propagation matrix $\mathcal{P}$,
\begin{align}
\mathcal{P}_{m,m+1} &= \begin{pmatrix}
e^{-ik_{z,1}d_{TI}}  \mathbbm{1}_{2}& 0 \\
0 & e^{i k_{z,1}d_{TI}}  \mathbbm{1}_{2} 
\end{pmatrix} , \label{eq-application-6}
\end{align}
one can obtain the transfer matrix for the structure in Fig.~\ref{figure-bilayer-fields-amplitudes}:
\begin{align}
\mathcal{T} = \mathcal{I}_0 \mathcal{P} \mathcal{I}_1 = 
\begin{pmatrix}
\mathcal{T}_{11} & \mathcal{T}_{12} \\
\mathcal{T}_{21} & \mathcal{T}_{22}
\end{pmatrix} . \label{eq-application-7}
\end{align}

The surface DPPMP modes in the TI/AFM bilayer satisfy the condition

\begin{align}
\det[\mathcal{T}_{11}] = 0 . \label{eq-application-8}
\end{align}

Since no external magnetic field is applied, the TE and TM modes remain independent of each other. As a result, the transfer matrix element $\mathcal{T}_{11}$ takes a diagonal form,
\begin{align}
\mathcal{T}_{11} = 
\begin{pmatrix}
T^{11}_{11} & 0 \\
0 & T^{22}_{22}
\end{pmatrix} . \label{eq-application-9}
\end{align}

The solutions of \eqref{eq-application-8} correspond to the conditions $T^{11}_{11} = 0$ and $T^{22}_{22} = 0$, which are associated with the TM and TE polarizations, respectively, of the electromagnetic wave incident on the structure. Since TE-polarized light cannot excite Dirac plasmons on the surface of the topological insulator (TI) material~\cite{ref49}, we restrict our analysis to the TM polarization. This case is determined by the condition $T^{11}_{11} = 0$, which leads to

\begin{widetext}

\begin{align}
&    \frac{e^{-i k_{z,1} d_{\mathrm{TI}}} k_{z,0} \alpha^2 \mu_0^2 \omega^2 - e^{i k_{z,1} d_{\mathrm{TI}}} k_{z,0} \alpha^2 \mu_0^2 \omega^2}{(k_x^2 + k_{z,0}^2) \mu_{yy}^{\mathrm{TI}} k_{z,1}}    +   \nonumber\\
& +\left[1 + \frac{k_{z,0}(k_x^2 + k_{z,1}^2)}{\mu_{yy}^{\mathrm{TI}} k_{z,1}(k_x^2 + k_{z,0}^2)} + \frac{\mu_0 \omega k_{z,0} \sigma_0}{(k_x^2 + k_{z,0}^2)}\right] 
\left[1 + \frac{\mu_{yy}^{\mathrm{TI}} k_{z,1}(k_x^2 + k_{z,2}^2)}{\mu_{yy}^{\mathrm{AFM}} k_{z,2}(k_x^2 + k_{z,1}^2)} +  \frac{\mu_0 \mu_{yy}^{\mathrm{TI}} \omega k_{z,1} \sigma_1}{(k_x^2 + k_{z,1}^2)}\right] e^{-i k_{z,1} d_{\mathrm{TI}}} + \nonumber \\
&+ \left[1 - \frac{k_{z,0}(k_x^2 + k_{z,1}^2)}{\mu_{yy}^{\mathrm{TI}} k_{z,1}(k_x^2 + k_{z,0}^2)} + \frac{\mu_0 \omega k_{z,0} \sigma_0}{(k_x^2 + k_{z,0}^2)}\right]  \, \left[1 - \frac{\mu_{yy}^{\mathrm{TI}} k_{z,1}(k_x^2 + k_{z,2}^2)}{\mu_{yy}^{\mathrm{AFM}} k_{z,2}(k_x^2 + k_{z,1}^2)}-\frac{\mu_0 \mu_{yy}^{\mathrm{TI}} \omega k_{z,1} \sigma_1}{(k_x^2 + k_{z,1}^2)}\right] e^{i k_{z,1} d_{\mathrm{TI}}}     = 0.  \label{eq-application-10}
\end{align}

\end{widetext}

The latter represents the general dispersion equation for a bi-isotropic bilayer structure, constituted of a bi-isotropic medium and AFM medium. By setting the bi-isotropic parameter equal to zero, one recovers the result reported in Ref.~\cite{ref55}. Unlike the conventional magnetoelectric coupling that often involves anisotropic tensors, the bi-isotropic coefficient $\alpha$ characterizes a scalar-type quantity that links electric and magnetic fields in an isotropic manner. The non-null $\alpha$ modifies the general dispersion relations, leading to measurable shifts in the dispersion of the surface Dirac plasmon--phonon--magnon polaritons (DPPMPs). Specifically, this scalar coupling introduces a cross-polarization effect that can alter the hybridization strength between the surface plasmons in the topological insulator and the magnetic excitations in the antiferromagnetic layer. As a result, one expects modifications in the spectral position, intensity, and possibly the linewidth of the DPPMP modes, particularly in the terahertz range, where these interactions are expected to be most prominent~\cite{ref55,ref49,ref61,ref62,ref63,ref65}.

Moreover, equation (\ref{eq-application-10}) is general and can be applied to a wide range of TI/AFM bilayer systems. It can be solved numerically to obtain the dispersion relation of surface polaritons in TI/AFM structures, provided the optical response functions and the thicknesses of the constituent layers are known. It is worth noting that, for $p$-polarized waves, the magnetic field of the electromagnetic wave lies along the $y$-direction. This means that an antiferromagnetic (AFM) magnetization aligned along the same direction would not influence the dispersion of the surface DPPMPs, since the corresponding component of the permeability tensor becomes unity $\mu_{yy}^{AFM} = 1$. Therefore, we consider the case where the AFM magnetization is aligned along the $x$-direction, i.e., perpendicular to the magnetic field of the EM wave, which results in a permeability tensor of the form:

\begin{align}
\mu_{yy}^{AFM} &= \mu^{AFM} = 1 + 4\pi \frac{2\gamma^2 H_a M}{\Omega_0^2 - \left(\omega + i/\tau_{mag}\right)^2} . \label{eq-application-11}
\end{align}

The solution of \eqref{eq-field-5} for the specified configuration yields the bulk modes in each region, which are given by:
\begin{align}
k_{z,0} &= \sqrt{\frac{\omega^{2}}{c^{2}}- k_x^{2}}, \label{eq-application-12} \\
k_{z,1} &= \sqrt{\frac{\omega^{2}}{c^{2}}\varepsilon^{TI} \mu^{TI}  - \mu_{0}^{2} \alpha^{2}\omega^{2}- k_x^{2}}, \label{eq-application-13} \\
k_{z,1} &= \sqrt{\frac{\omega^{2}}{c^{2}} \varepsilon^{AFM} \mu^{AFM} - k_x^{2}}. \label{eq-application-14}
\end{align}

In the considered configuration, we focus on a ribbon geometry where both the topological insulator (TI) and the antiferromagnetic (AFM) materials are confined along the $x$-direction. Under such confinement, the in-plane wavevector component can be approximated as $k_x = q \approx \frac{\pi}{W}$, where $W$ is the width of the TI/AFM ribbon along the $x$-axis. This quantization condition arises from the boundary constraints imposed by the finite lateral size of the heterostructure. It is important to note that, in this analysis, we assume that the TI material does not possess any intrinsic magnetic ordering. As a consequence, the magnetic response of the TI is trivial, and it does not contribute to the permeability tensor. Therefore, its magnetic permeability can be taken as unity and isotropic, leading to the $\mu^{TI}  = 1$.

Substituting the relations (\ref{eq-application-11})–(\ref{eq-application-14}) into \eqref{eq-application-10}, we finally obtain

\begin{widetext}

\begin{align}
G^{-1} &= \frac{k_{z,0}}{k_{z,1}} \alpha^{2} \mu_{0}^{2} c^{2} 
\left( e^{-i k_{z,1} d_{TI}} - e^{i k_{z,1} d_{TI}} \right) + \left( 1 + \frac{k_{z,0} \varepsilon^{TI}}{k_{z,1}} + \frac{k_{z,0} \sigma_{0}}{\varepsilon_{0} \omega} \right)
\left( 1 + \frac{\varepsilon^{AFM} k_{z,1}}{\varepsilon^{TI} k_{z,2}} + \frac{k_{z,1} \sigma_{1}}{\varepsilon_{0} \varepsilon^{TI} \omega} \right)
e^{-i k_{z,1} d_{TI}} \nonumber \\
&\quad + \left( 1 - \frac{k_{z,0} \varepsilon^{TI}}{k_{z,1}} + \frac{k_{z,0} \sigma_{0}}{\varepsilon_{0} \omega} \right)
\left( 1 - \frac{\varepsilon^{AFM} k_{z,1}}{\varepsilon^{TI} k_{z,2}} - \frac{k_{z,1} \sigma_{1}}{\varepsilon_{0} \varepsilon^{TI} \omega} \right)
e^{i k_{z,1} d_{TI}} = 0 . \label{eq-application-16}
\end{align}

\end{widetext}

\subsection{\label{section-cr2o3}Material parameters for specific TI/AFM systems}

To explore the surface DPPMP behavior in selected TI/AFM heterostructures, the analysis is restricted to a single topological insulator candidate, Bi$_2$Se$_3$, whose bulk dielectric response in the far-infrared regime of interest is provided in Ref.~\cite{ref49}, namely,
\begin{align}
\varepsilon^{TI} = \varepsilon_{\infty} 
+ \frac{S_{\alpha}^2}{\omega_{\alpha}^2 - \omega^2 - i\omega\Gamma_{\alpha}}
+ \frac{S_{\beta}^2}{\omega_{\beta}^2 - \omega^2 - i\omega\Gamma_{\beta}} , \label{eq-application-17}
\end{align}
here, $\omega_{j}$, $\Gamma_{j}$, and $S_{j}$ represent the resonance frequency, damping rate, and oscillator strength corresponding to the Lorentzian terms associated with the $\alpha$ ($j = \alpha$) and $\beta$ ($j = \beta$) phonon modes of the TI thin film. The numerical values of all parameters associated with Bi$_2$Se$_3$, along with those of other representative topological insulators for reference, are summarized in Table~\ref{table1}.
\begin{figure}[h]
\begin{centering}
\includegraphics[scale=0.55]{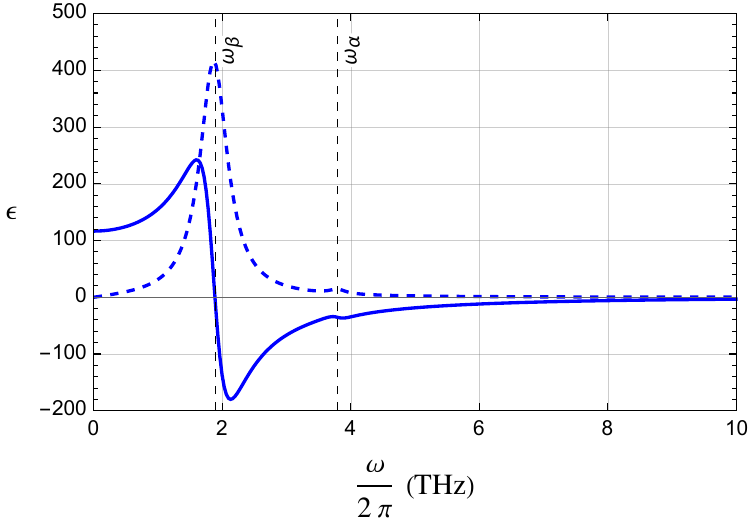}
\par\end{centering}
\caption{\small{\label{figure-permittiviy-Bi2Se3}}Frequency-dependent dielectric function of Bi$_{2}$Se$_{3}$ given in \eqref{eq-application-17}. The solid (dashed) line indicates the real (imaginary) parts of $\epsilon(\omega)$. Here, we have used the values of Tab.~\ref{table1}.}
\end{figure}

The surface states of these topological insulators support two-dimensional, spin-polarized Dirac plasmons, which effectively act as a conductive electron sheet. The optical response of this sheet is characterized by its surface conductivity, expressed as
\begin{align}
\sigma = \frac{e^{2} E_{F}}{4 \pi \hbar^{2} \left( i\omega - \tau_{DP}^{-1} \right)} . \label{eq-application-18}
\end{align}
Here, $E_{F}$ denotes the Fermi energy associated with the topological surface states, $\tau_{DP}$ represents the relaxation time of the Dirac plasmon mode, and $e$ is the elementary charge. It is worth mentioning that the emergence of hybridized excitations at the interface between the topological insulator (TI) and the adjacent medium, here, the antiferromagnet (AFM), can modify the interfacial carrier density compared to that of an isolated TI surface, as suggested in Refs.~\cite{ref49,ref59}. Nonetheless, in the present treatment, we disregard such interfacial renormalization effects and assume an identical surface conductivity for both the bare TI surface and the TI/AFM interface. Accordingly, we adopt the notation $\sigma_{0} \equiv \sigma_{1} \equiv \sigma$, with $\sigma$ defined in \eqref{eq-application-18}.

It is worth mentioning that, in the long-wavelength limit ($k_{x} \cdot d_{TI} \ll 1$), an analytical expression for the surface plasmon--phonon polariton in the thin TI film was derived in Ref.~\cite{ref59},
\begin{align}
\omega^{2}_{TI+} &= \frac{v_{F} \sqrt{2 \pi n_{2D}} e^{2}}{\varepsilon_{0} h} 
\frac{k_{x}}{\varepsilon_{\text{top}} + \varepsilon_{\text{bot}} + k_{x} d_{TI} \varepsilon_{TI}} , \label{eq-application-19} \\
\omega^{2}_{TI-} &= 
\frac{2 \varepsilon_{0} \varepsilon_{TI} h v_{F} + e^{2} \sqrt{2 \pi n_{2D} d_{TI}}}
{\sqrt{4 \pi \varepsilon_{0} \varepsilon_{TI}^{2} h^{2} v_{F}^{2} + 2 \varepsilon_{0} \varepsilon_{TI} e^{2} \sqrt{2 \pi n_{2D} d_{TI}}}} 
\, k_{x}^{2} , \label{eq-application-20}
\end{align}

The subscripts TI$^{+}$ and TI$^{-}$ represent the optical and acoustic branches, respectively. Here, $\upsilon_{F}$ denotes the Fermi velocity of the Dirac plasmons in the topological insulator. The parameter $n_{2D}$ corresponds to the sheet carrier density of the entire TI thin film, accounting for the contributions from both surfaces. The quantities $\varepsilon_{\text{top}}$, $\varepsilon_{\text{bot}}$, and $\varepsilon_{TI}$ denote the permittivities of the top dielectric, bottom dielectric, and the TI layer, respectively. The variable $k_{x}$ corresponds to the in-plane wavevector, while $d_{TI}$ denotes the thickness of the TI film. This work concentrates on the optical mode of the surface plasmon polariton in the TI, since it is the only branch that can be efficiently excited in standard optical measurements. The acoustic mode is excluded because it does not contribute to the optical dipole matrix element~\cite{ref70}.

\begin{widetext}

\begin{table}[H]
\caption{The TI parameters used in this work from Ref.~\cite{ref73}.}
\centering
\begin{tabular}{l C{1.8cm} C{1.8cm} C{1.8cm} C{1.8cm} C{1.8cm} C{1.8cm} C{1.8cm} C{1.8cm}}
\hline\hline \\[-1.5ex]
Materials & $\varepsilon_{\infty}$ & $S_{\alpha}~(\mathrm{cm}^{-1}) $ & $\omega_{\alpha}~(\mathrm{cm}^{-1}) $ & $\Gamma_{\alpha}~(\mathrm{cm}^{-1}) $ & $S_{\beta}~(\mathrm{cm}^{-1}) $ & $\omega_{\beta}~(\mathrm{cm}^{-1}) $ & $\Gamma_{\beta}~(\mathrm{cm}^{-1}) $ \\
\hline\\[-2ex]
Bi$_2$Se$_3$ & 1  & 675.9  & 63.03 & 17.5 & 100 & 126.94 & 10 \\[0.5ex]
Bi$_2$Te$_3$ & 85 & 716    & 50    & 10   & 116 & 95     & 15 \\[0.5ex]
Sb$_2$Te$_3$ & 51 & 1498.0 & 67.3  & 10   & NA & NA & NA \\
\hline\hline
\end{tabular}
\label{table1} 
\end{table}

\end{widetext}

In this work, we focus on investigating two candidate antiferromagnetic (AFM) materials, Cr$_2$O$_3$ and FeF$_2$, both of which support magnon excitations in the terahertz (THz) frequency range~\cite{ref23,ref72}. Notably, Cr$_2$O$_3$ exhibits a N\'eel temperature around 308~K, enabling its experimental study under ambient conditions~\cite{ref28}. In contrast, FeF$_2$ and MnF$_2$ must be examined at cryogenic temperatures. For our calculations, we assume all AFM samples remain below their respective N\'eel temperatures. The frequency-dependent magnetic permeability of these AFMs is modeled using Eq.~(33), and the relevant magnetic parameters, including the characteristic magnon frequencies, are summarized in Table~\ref{table2}. In addition to Cr$_2$O$_3$ and FeF$_2$, Table~\ref{table2} also includes other representative AFM materials for comparison purposes. Regarding the dielectric properties, we adopt a constant relative permittivity value of $\varepsilon_{AFM} = 5$ for all AFM materials, consistent with the assumption employed in Ref.~\cite{ref55}.

\begin{widetext}


\begin{table}[H]
\caption{The parameters for AFM materials used in this paper.}
\centering
\begin{minipage}{\textwidth}
\centering
\begin{tabular}{l C{2cm} C{2cm} C{2cm} C{2cm} C{2cm} C{2cm} C{2cm}}
\hline\hline
Materials & $H_{a}$ (Oe) & $H_{e}$ (Oe) & $M$ (G) & $\Omega_{0}$ (THz) & $\tau_{\mathrm{mag}}$ (ns) & Land\'e Factor & $T_{\mathrm{Neel}}$ (K) \\[0.5ex]
\hline\\[-2ex]
Cr$_2$O$_3$~\cite{ref24,ref25,ref72} & $7.2\times10^{2}$ & $2.45\times10^{6}$ & 590 & 0.17 & $\sim$0.14 & $\sim$2.0 & 308 \\[0.5ex]
NiO~\cite{ref23,ref26,ref82}        & $6.4\times10^{3}$ & $9.7\times10^{6}$ & 400 & 1.01 & 0.0175$^{\mathrm{b}}$ & 2.05 & 523 \\[0.5ex]
MnF$_2$~\cite{ref23,ref85}          & $8\times10^{3}$ & $5.33\times10^{5}$ & 592 & 0.26 & 7.58 & 2.0 & 67 \\[0.5ex]
FeF$_2$~\cite{ref86}                 & $2\times10^{5}$ & $5.4\times10^{5}$  & 560 & 1.62 & 0.11 & 2.25 & 78 \\
\hline\hline
\end{tabular}

\vspace{2pt}
\raggedright
\footnotesize
$^{\dagger}$ $Mn^{2+}$/Mn$^{3+}$ in metallic AFM typically exhibits $g$ near to 2. \\
$^{\mathrm{b}}$ The value is selected such that it corresponds to the width of the resonance line of 18~GHz. \\
$^{\mathrm{c}}$ The value is selected such that it corresponds to the width of the resonance line of ~100 GHz. \;
\end{minipage}
\label{table2}
\end{table}

\end{widetext}

\subsection{\label{section-fef2}Formation of surface plasmon-magnon polaritons in TI/AFM structures}

\begin{figure}[t]
\centering
\subfloat{\includegraphics[width=0.64\columnwidth]{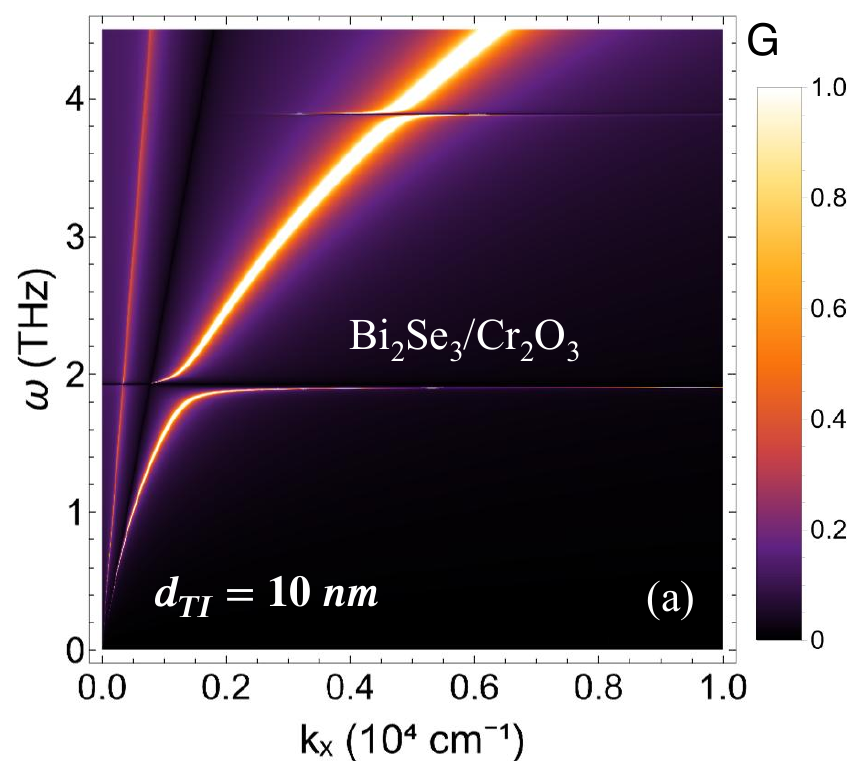}
\label{figure-coupling-Cr2O3-10-nm-A}}
\par%
\subfloat{\includegraphics[width=0.64\columnwidth]{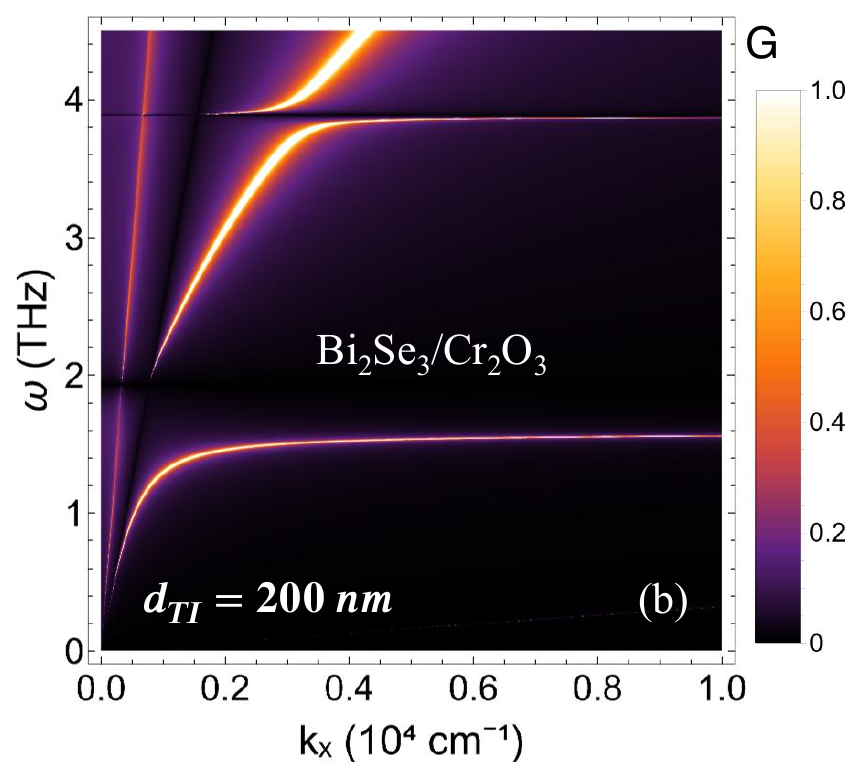}
\label{figure-coupling-Cr2O3-200-nm-B}}
\caption{\label{figure-coupling-Cr2O3}\small Surface DPPMP dispersion in Bi$_2$Se$_3$/Cr$_2$O$_3$ structure calculated using \eqref{eq-application-16} with the Fermi energy of the Dirac plasmon on its surface $E_{F} = 1~\mathrm{eV}$ and the thickness of TI layer (a) $d_{TI} = 10~\mathrm{nm}$ and (b) $d_{TI} = 200~\mathrm{nm}$.
}
\end{figure}

To illustrate the dispersion relations of surface plasmon-phonons in a Bi$_2$Se$_3$ layer interacting with a magnon in an AFM, we first select chromium(III) oxide (Cr$_2$O$_3$) as the antiferromagnetic medium due to its well-established collinear AFM order, high uniaxial magnetic anisotropy, and a N\'eel temperature near room temperature ($\approx 307$~K), which ensures experimental viability~\cite{ref74,ref75,ref76,ref72}. The strong magnetic anisotropy along the trigonal axis leads to a well-defined magnon gap in the terahertz range, making Cr$_2$O$_3$ an ideal platform for exploring spin--charge--lattice coupling in topological-antiferromagnetic heterostructures. Figure~\ref{figure-coupling-Cr2O3} presents the numerical evaluation of the function $G$, defined in Eq.~(\ref{eq-application-16}), as a function of frequency $\omega$ and the in-plane wavevector $k_{x}$.  We investigate Bi$_2$Se$_3$/Cr$_2$O$_3$ heterostructures using two typical topological insulator thicknesses commonly employed in thin-film devices~\cite{ref78,ref79,ref80,ref81}: 10~nm and 200~nm, shown in panels (a) and (b), respectively. The color in Fig.~\ref{figure-coupling-Cr2O3} represents the magnitude of the function $G$ whose maxima reveal the dispersion of the surface DPPMP.

In Fig.~\ref{figure-coupling-Cr2O3}\subref{figure-coupling-Cr2O3-10-nm-A}, for a relatively thin Bi$_2$O$_3$ layer (10~nm), we observe the emergence of surface DPPPs (Dirac plasmon--phonon polaritons) through distinct anticrossings near 2~THz and 4~THz. The feature at 2~THz arises from the hybridization of the TI plasmon with the $\alpha$ phonon mode of Bi$_2$O$_3$, while the 4~THz anticrossing is associated with coupling to the $\beta$ phonon mode. The magnon mode of Cr$_2$O$_3$ lies below $\sim 0.2$~THz and does not significantly interact with the plasmon branch in this frequency range. As we will discuss in the following sections, this result contrasts with the behavior observed for the Bi$_2$O$_3$/FeF$_3$ heterostructure, where the magnon frequency lies close to the plasmon resonance around 1.59~THz. This near-degeneracy leads to the emergence of a third anticrossing feature in the dispersion spectrum.

As depicted in Fig.~\ref{figure-coupling-Cr2O3}\subref{figure-coupling-Cr2O3-200-nm-B}, when the thickness of the Bi$_2$Se$_3$ layer is increased to $d_{TI} = 200$~nm, one notices that both contributions from $\alpha$ and $\beta$ phonons still hold in the dispersion, but the signature of the interaction between the Bi$_2$Se$_3$ and Cr$_2$O$_3$ layer at 2~THz completely disappears. This behavior arises from the thickness-dependent evolution of the surface plasmon--phonon polariton modes in the bare topological insulator, as governed by \eqref{eq-application-16}. As the Bi$_2$Se$_3$ film becomes thicker, the upper polaritonic branch experiences a notable blueshift, moving beyond 2~THz. This shift is attributed to the negative real part of the TI's dielectric function in this frequency range (see Fig.~\ref{figure-permittiviy-Bi2Se3}). In contrast, the lower branch, situated just below 2~THz, undergoes a redshift, consistent with the positive permittivity exhibited by Bi$_2$Se$_3$ in that spectral domain. The downward shift of the lower mode displaces it away from the plasmon resonance at 2~THz, effectively suppressing the mode degeneracy responsible for hybridization. As a result, the previously observed Bi$_2$Se$_3$/Cr$_2$O$_3$ anticrossing associated with the plasmon--$\alpha$ phonon hybrid mode disappears in Fig.~\ref{figure-coupling-Cr2O3}\subref{figure-coupling-Cr2O3-200-nm-B}. A similar behavior was observed for the other AFM materials listed in Table~\ref{table2} (not shown here).




On the other hand, increasing the TI thickness to $d_{\mathrm{TI}} = 200~\mathrm{nm}$ enhances the coupling strength between the surface plasmon mode and the $\beta$ optical phonon of the topological insulator. This enhancement is evidenced by a more pronounced anticrossing near $4~\mathrm{THz}$, where the upper and lower polaritonic branches exhibit a clear spectral separation. The increased field confinement within the thicker Bi$_2$O$_3$ layer strengthens the interaction between the plasmonic mode and the $\beta$ phonon, resulting in a broader hybridization gap. This behavior contrasts with the thinner-film case, in which the coupling is weaker and the anticrossing less defined. The observation reinforces the crucial role of geometrical parameters, such as TI thickness, in modulating the plasmon--phonon interaction and tuning the spectral response of Dirac plasmon--phonon polaritons (DPPPs) in TI/AFM heterostructures.
\begin{widetext}

\begin{figure}[h]
\centering
\subfloat{\includegraphics[width=0.29\textwidth]{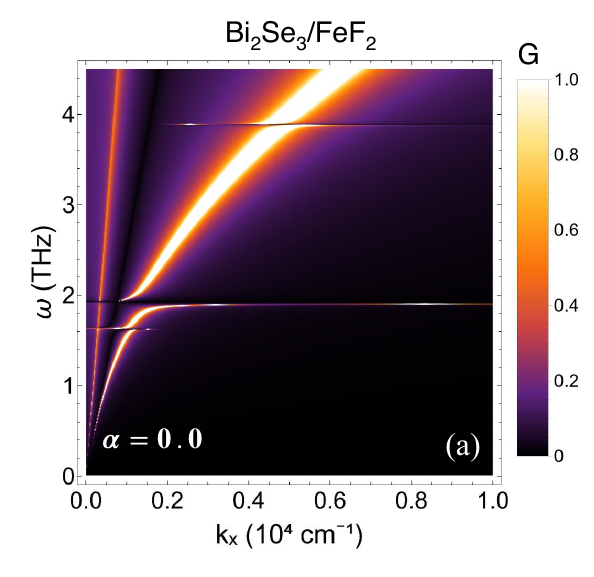}
\label{figure-coupling-FeF2-alpha-A}}
\hfill
\subfloat{\includegraphics[width=0.29\textwidth]{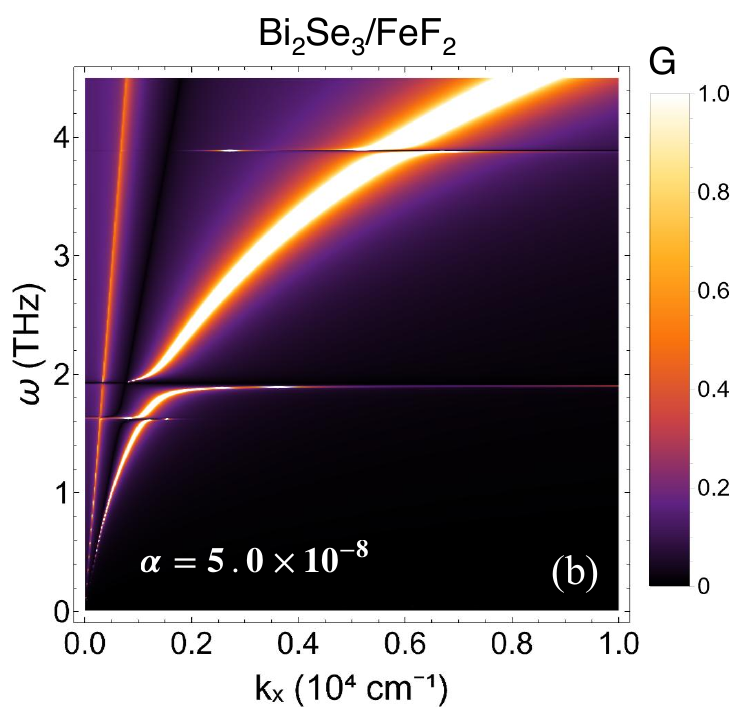}
\label{figure-coupling-FeF2-alpha-B}}
\hfill
\subfloat{\includegraphics[width=0.29\textwidth]{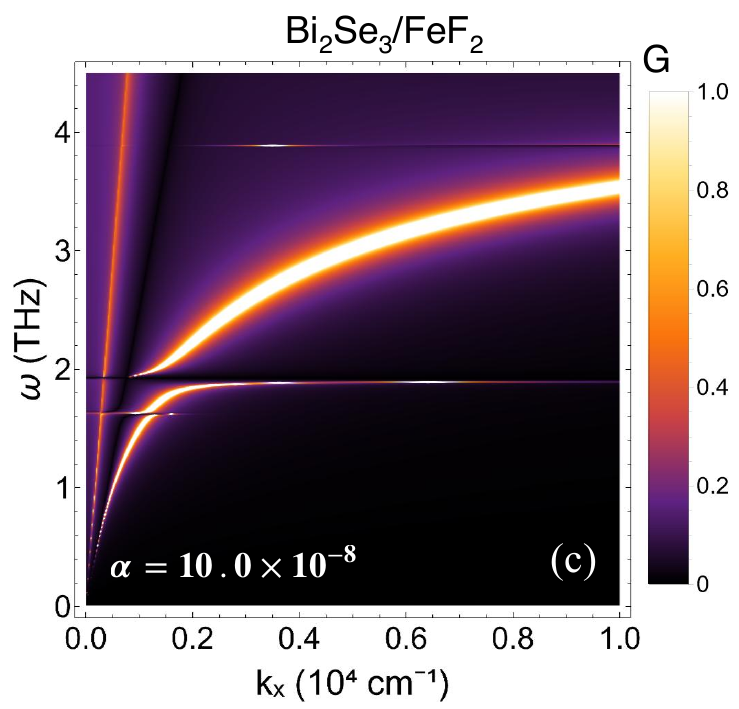}
\label{figure-coupling-FeF2-alpha-C}}
\captionof{figure}{\label{figure-coupling-FeF2-alpha-all-cases}\small
Dispersipon of DPPMP in a Bi$_2$Se$_3$/FeF$_2$ structure, calculated using Eq.~(\ref{eq-application-16}), with the Fermi energy of the Dirac plasmon on the surface set to $E_{F} = 1~\mathrm{eV}$, and the topological insulator (TI) layer thickness $d_{TI} = 10~\mathrm{nm}$.}
\end{figure}

\end{widetext}

\subsection{\label{section-influence-bi-isotropic-on-coupling-strength}Influence of the bi-isotropic parameter on the coupling strength}

We now examine how the bi-isotropic coupling parameter $\alpha$ modifies the interaction strength between surface plasmon modes and other excitations. By introducing $\alpha$ parameter, its impact on the hybridization behavior and the emergence or suppression of anticrossing features in the dispersion relations of DPPMPs is assessed. We begin by analyzing the dispersion relations of surface plasmon--phonon modes in a Bi$_2$Se$_3$ topological insulator (TI) layer coupled to magnons in a FeF$_2$ antiferromagnet (AFM). Figure~\ref{figure-coupling-FeF2-alpha-all-cases} presents the function $G$, as defined by Eq.~(\ref{eq-application-16}), plotted as a function of the frequency $\omega$ and the in-plane wavevector $k_{x}$. We investigate the effects of the bi-isotropic parameter $\alpha$, considering three cases: (a) absence of coupling ($\alpha = 0$), (b) $\alpha = 5\times 10^{-8}$, and (c) $\alpha = 10\times 10^{-8}$. In all scenarios, the Fermi energy of the Dirac plasmon is fixed at $E_{F} = 1~\mathrm{eV}$. In the absence of bi-isotropic coupling, distinct anticrossings emerge around 1.5~THz, 2~THz, and 4~THz, which indicate resonant interactions where the Dirac plasmon mode becomes degenerate with, respectively, the magnon excitation in the FeF$_2$ layer, and the $\alpha$ and $\beta$ optical phonons in the Bi$_2$Se$_3$ film. Analogous behavior has been observed for Bi$_2$Se$_3$/FeF$_2$ bilayers \cite{ref55}.

\begin{widetext}

\begin{figure}[h]
\centering
\subfloat{\includegraphics[width=0.3\textwidth]{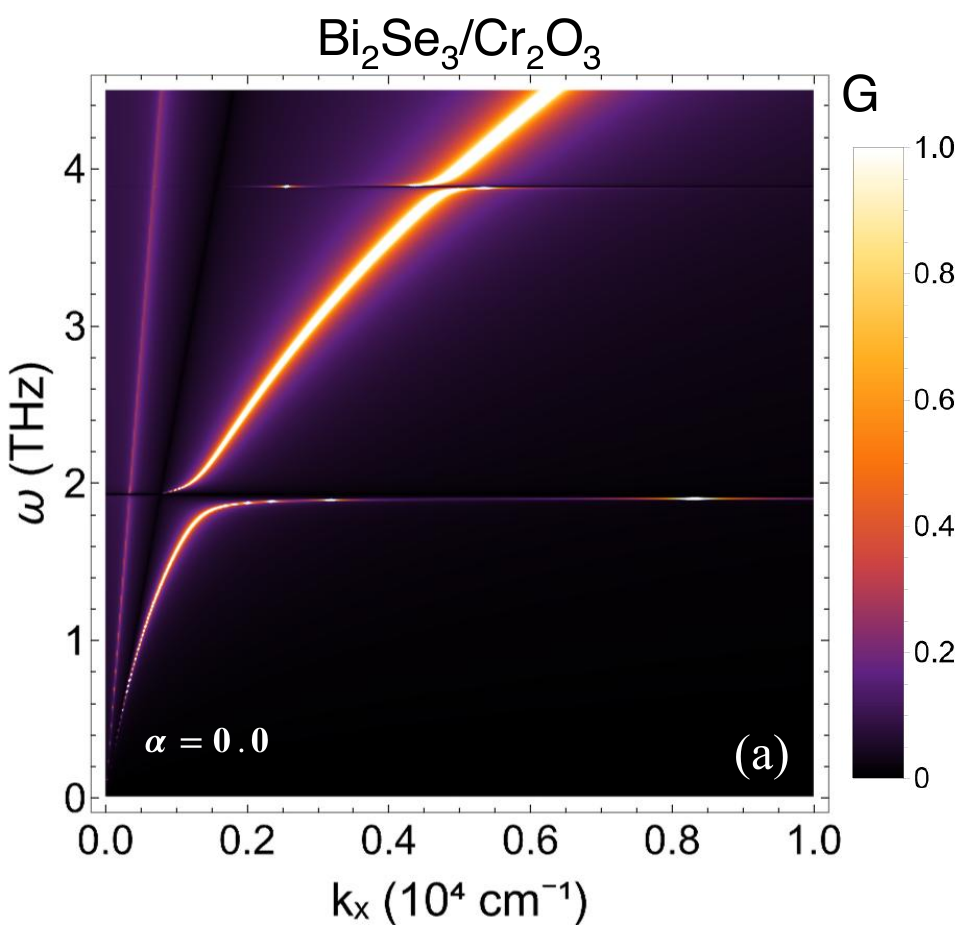}
\label{figure-coupling-Cr2O3-alpha-A}}
\hfill
\subfloat{\includegraphics[width=0.3\textwidth]{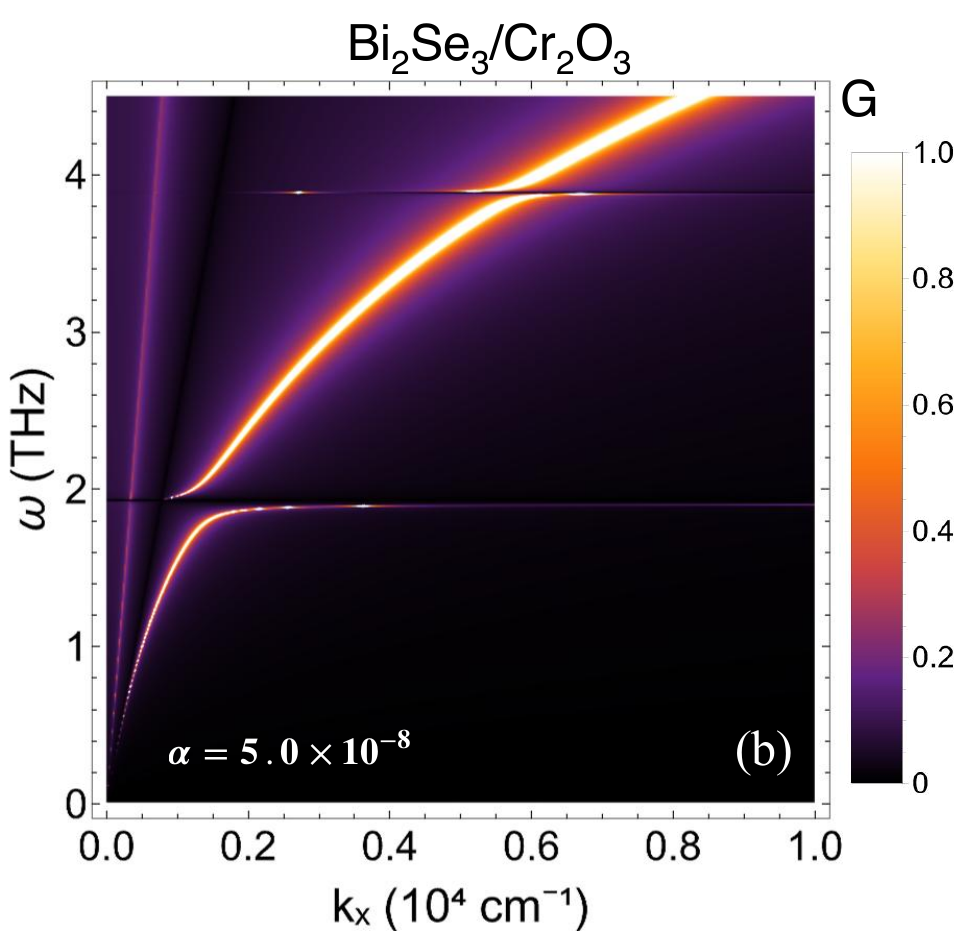}
\label{figure-coupling-Cr2O3-alpha-B}}
\hfill
\subfloat{\includegraphics[width=0.3\textwidth]{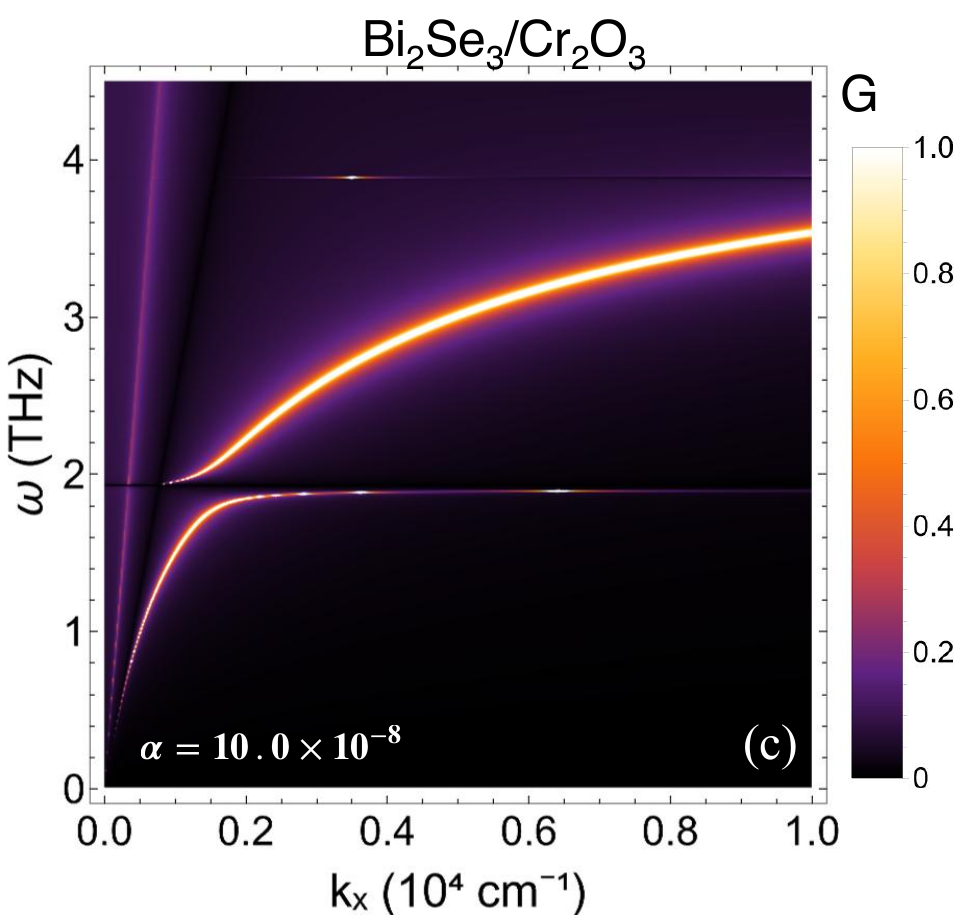}
\label{figure-coupling-Cr2O3-alpha-C}}
\captionof{figure}{\label{figure-coupling-Cr2O3-alpha-all-cases}\small Dispersion of DPPMPs in a Bi$_2$Se$_3$/FeF$_2$ heterostructure for different values of the bi-isotropic parameter $\alpha$: (a) $\alpha = 0$, (b) $\alpha = 10\times 10^{-8}$, and (c) $\alpha = 15\times 10^{-8}$. Here, the Fermi energy of the Dirac plasmon on the surface is set to $E_{F} = 1~\mathrm{eV}$, and the topological insulator (TI) layer thickness $d_{TI} = 10~\mathrm{nm}$.}
\end{figure}

\end{widetext}

As shown in Fig.~\ref{figure-coupling-FeF2-alpha-all-cases}\subref{figure-coupling-FeF2-alpha-B}, when the bi-isotropic parameter assumes non-null values, the $\alpha$ and $\beta$ phonon contributions remain visible in the dispersion. However, the interaction between the Bi$_2$Se$_3$ and FeF$_2$ layers leads to a redshift of the upper polaritonic branch near 4~THz as the bi-isotropic term $\alpha$ increases. In Fig.~\ref{figure-coupling-FeF2-alpha-all-cases}\subref{figure-coupling-FeF2-alpha-C}, corresponding to the strong bi-isotropic parameter regime $\alpha = 10\times10^{-8}$, the dispersion of the surface Dirac plasmon--phonon--magnon polariton (DPPMP) exhibits notable modifications compared to the uncoupled and moderately coupled cases, with strong redshifts of the upper branch due to the negative real part of its permittivity in this domain. Although the $\alpha$ and $\beta$ phonon modes of Bi$_2$Se$_3$ remain visible, the anticrossing associated with the hybridization between the Dirac plasmon and the $\beta$ phonon (plasmon--phonon polariton) becomes strongly suppressed. The disappearance of the anticrossing feature in this region suggests that the hybrid interaction between the Bi$_2$Se$_3$ and FeF$_2$ layers at this frequency is significantly weakened or inhibited. This behavior may indicate a saturation effect in the coupling mechanism or a detuning between the interacting modes as the bi-isotropic term $\alpha$ increases. A similar behavior was also observed in the Bi$_2$Se$_3$/Cr$_2$O$_3$ heterostructure, as shown in Fig.~\ref{figure-coupling-Cr2O3-alpha-all-cases}(\subref*{figure-coupling-Cr2O3-alpha-A}--\subref*{figure-coupling-Cr2O3-alpha-C})
, where the increase in $\alpha$ leads to a comparable redshift of the upper polaritonic branch and suppression of the anticrossing associated with the plasmon--phonon coupling.

Conceptually, the presence of $\alpha$ alters the electromagnetic interaction at the TI/AFM interface, leading to a modified dispersion relation. Although the $\beta$ phonon is intrinsic to the topological insulator, the bi-isotropic parameter still impacts its hybridization with the plasmonic branch by redistributing field confinement across the heterostructure. As $\alpha$ increases, the upper polaritonic branch undergoes a redshift and eventually moves below the 4.0~THz region, where it previously exhibited a strong anticrossing with the optical phonon mode of the topological insulator. This behavior indicates that the bi-isotropic interaction not only perturbs the hybridization conditions but also leads to a suppression of the polaritonic signature associated with the $\beta$ phonon. The resulting state reflects a reconfiguration of the surface mode, possibly indicating a breakdown of the original surface-bound resonance rather than the formation of a conventional surface--bulk hybrid excitation.

\begin{widetext}

\begin{figure}[h]
\centering
\subfloat{\includegraphics[width=0.32\textwidth]{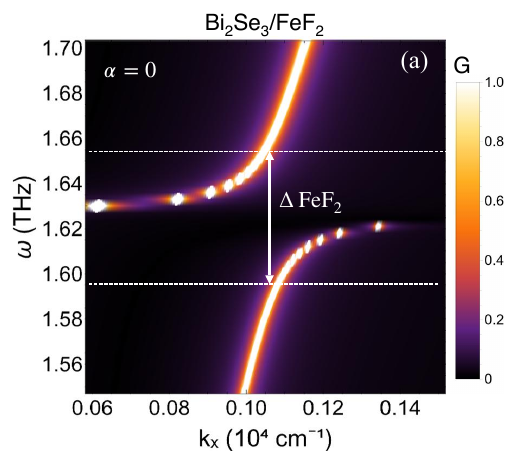}
\label{figure-coupling-strength-FeF2-alpha-A}}
\hfill
\subfloat{\includegraphics[width=0.32\textwidth]{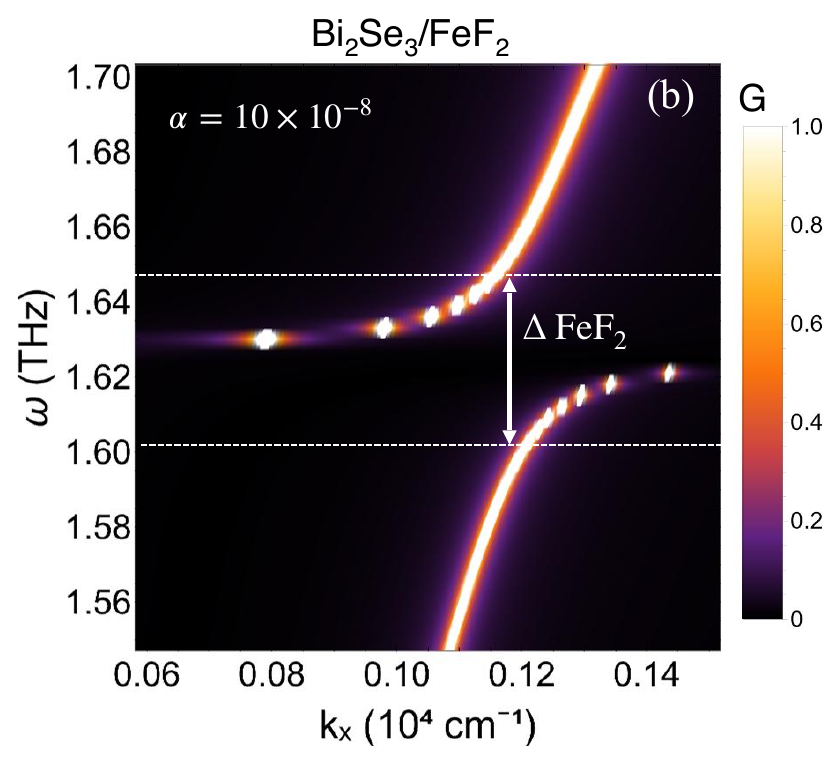}
\label{figure-coupling-strength-FeF2-alpha-B}}
\hfill
\subfloat{\includegraphics[width=0.32\textwidth]{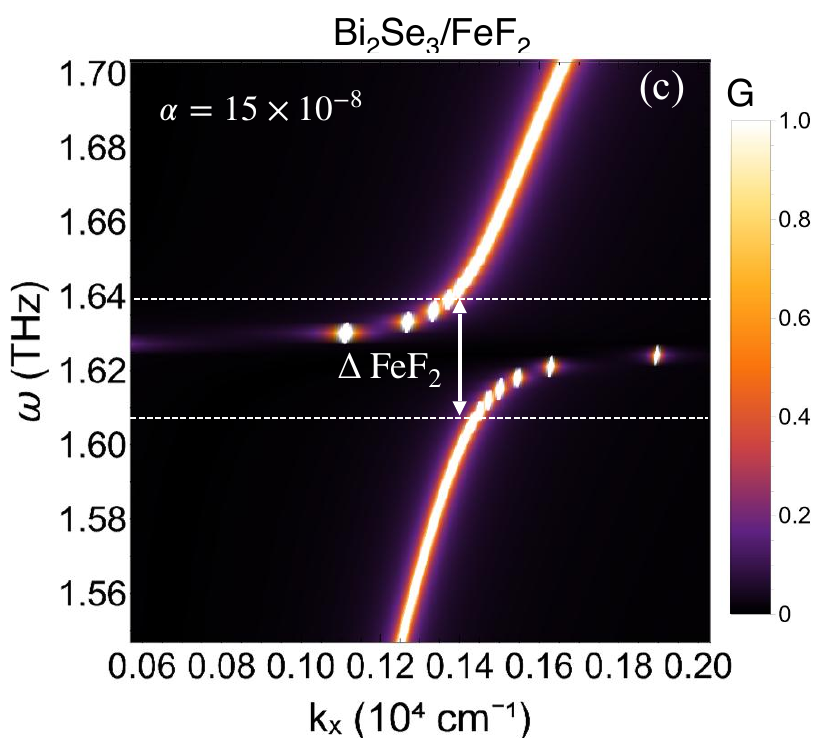}
\label{figure-coupling-strength-FeF2-alpha-C}}
\captionof{figure}{\label{figure-coupling-strength-FeF2-alpha-all-cases-separadas}\small Dispersion of Dirac plasmon--phonon--magnon polaritons (DPPMPs) in a Bi$_2$Se$_3$/FeF$_2$ heterostructure for different values of the bi-isotropic coupling parameter $\alpha$: (a) $\alpha = 0$, (b) $\alpha = 10\times 10^{-8}$, and (c) $\alpha = 15\times 10^{-8}$. The colormap represents the magnitude of the spectral function $F$, whose maxima trace the hybrid mode dispersions. The white dashed lines mark the upper and lower branches of the hybrid modes near the magnon resonance frequency. The splitting between these branches, denoted by $\Delta_{\mathrm{FeF_2}}$, quantifies the coupling strength between the Dirac surface plasmons in the topological insulator and the magnon modes in the antiferromagnet.}
\end{figure}

\end{widetext}

Our previous analyses have primarily focused on the effect of the bi-isotropic parameter $\alpha$ on the upper hybrid mode. However, a more comprehensive investigation reveals subtle yet physically relevant modifications in the dispersion of Dirac plasmon--phonon--magnon polaritons (DPPMPs) near the magnon resonance, which exhibit a pronounced dependence on $\alpha$. As shown in Fig.~\ref{figure-coupling-strength-FeF2-alpha-all-cases-separadas}(\subref*{figure-coupling-strength-FeF2-alpha-A}--\subref*{figure-coupling-strength-FeF2-alpha-C}), the coupling strength, which is defined as the frequency splitting between the upper and lower branches in the vicinity of the magnon frequency and around the anticrossing wave vector, decreases monotonically with increasing $\alpha$. Specifically, the mode separation is reduced from $\Delta \approx 0.06$~THz in the absence of bi-isotropic coupling to $\Delta \approx 0.03$~THz for $\alpha = 15\times 10^{-8}$, indicating that the bi-isotropic parameter tend to suppress the hybridization strength between the coupled modes. Thus, the pronounced sensitivity of the coupling strength to $\alpha$ underscores the central role of topological electromagnetic coupling in shaping the polariton spectrum. These findings open new pathways for the active control of light--matter interactions in engineered TI/AFM platforms, particularly within the technologically relevant terahertz regime.

\subsection{\label{section-impact-bi-isotropic}Impact of the bi-isotropic parameter and Fermi energy on the coupling strength}

\begin{widetext}

\begin{figure}[h]
\centering
\subfloat{\includegraphics[width=0.32\textwidth]{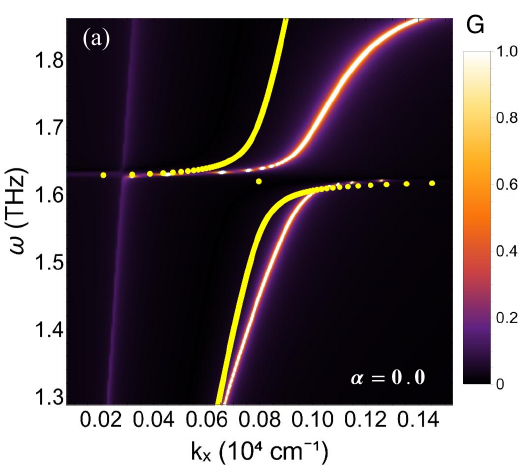}
\label{figure-blueshift-redshift-alpha-A}}
\hfill
\subfloat{\includegraphics[width=0.32\textwidth]{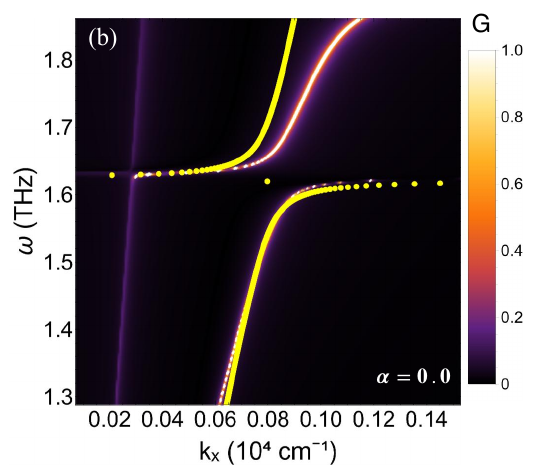}
\label{figure-blueshift-redshift-alpha-B}}
\hfill
\subfloat{\includegraphics[width=0.32\textwidth]{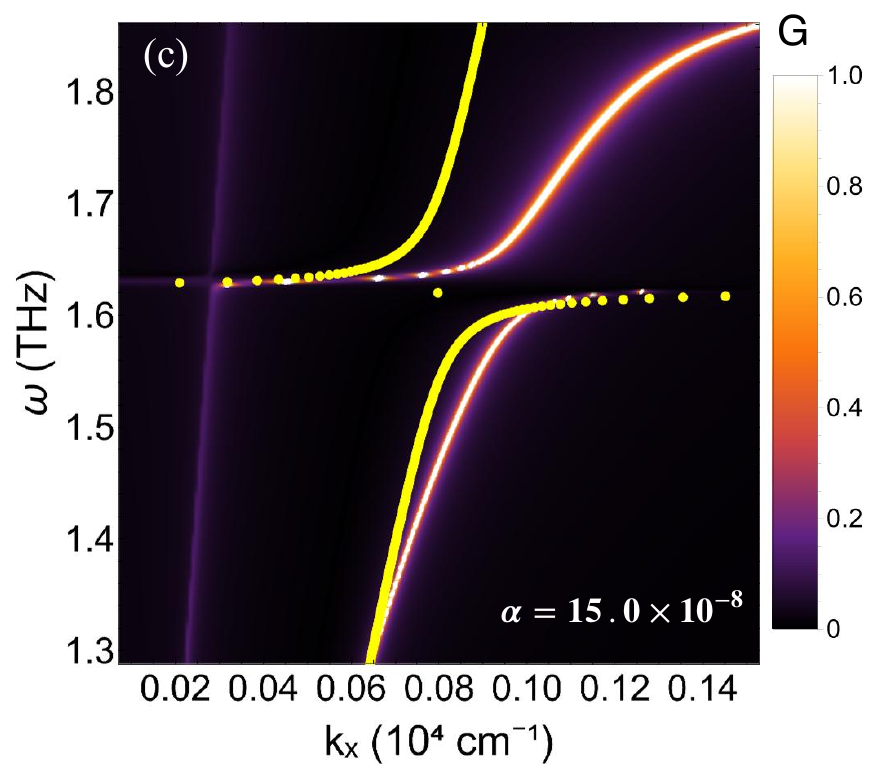}
\label{figure-blueshift-redshift-alpha-C}}
\captionof{figure}{\label{figure-blueshift-redshift-alpha-all-cases-separadas}\small Dispersion of Dirac plasmon--phonon--magnon polaritons (DPPMPs) in a Bi$_2$Se$_3$/FeF$_2$ heterostructure for different values of the bi-isotropic coupling parameter $\alpha$: (a) $\alpha = 0$, (b) $\alpha = 10\times 10^{-8}$, and (c) $\alpha = 15\times 10^{-8}$. The colormap represents the magnitude of the spectral function $F$, whose maxima trace the hybrid mode dispersions. The white dashed lines mark the upper and lower branches of the hybrid modes near the magnon resonance frequency. The splitting between these branches, denoted $\Delta_{\mathrm{FeF_2}}$, quantifies the coupling strength between the Dirac surface plasmons in the topological insulator and the magnon modes in the antiferromagnet.}
\end{figure}

\end{widetext}

We proceed by analyzing how variations in the Fermi level of the surface states in the topological insulator (TI)
layer affects the coupling behavior between the TI and the antiferromagnetic (AFM) medium. Figure~\ref{figure-blueshift-redshift-alpha-all-cases-separadas} presents the dispersion of surface DPPMP modes in a Bi$_2$Se$_3$/FeF$_2$ bilayer as the Fermi energy is increased from $1.0$ to $1.5$~eV. As the Fermi level rises, all polaritonic branches undergo a noticeable blueshift. This behavior can be attributed to the enhanced population of surface electrons contributing to the collective plasmonic mode, thereby shifting the dispersion to higher frequencies in agreement with \eqref{eq-application-19}.

One key implication of this blueshift is that increasing the Fermi level effectively brings the Dirac plasmon--phonon polariton (DPPP) closer to resonance with the magnon--polariton mode in the AFM, leading to a stronger coupling. To illustrate this effect, Fig.~\ref{figure-blueshift-redshift-alpha-all-cases-separadas} displays the surface DPPMP dispersion in the Bi$_2$Se$_3$/FeF$_2$ system for three representative scenarios: (a) $E_{F} = 1.2~\mathrm{eV}$, with $\alpha = 0$, (b) $E_{F} = 1.5~\mathrm{eV}$ with $\alpha = 0$, and (c) $E_{F} = 1.5~\mathrm{eV}$ with $\alpha = 15\times 10^{-8}$.

In all plots, the dispersion of the bulk magnon polariton in FeF$_2$ is shown in yellow and follows the relation $k^{2} =\varepsilon_{\mathrm{AFM}} \mu_{\mathrm{AFM}} \omega^{2}/c^{2}$. For the lower Fermi energy case [Fig.~\ref{figure-blueshift-redshift-alpha-all-cases-separadas}\subref{figure-blueshift-redshift-alpha-A}], the Dirac plasmon--phonon--polariton crosses the magnon resonance frequency ($\omega_{0} = 1.62~\mathrm{THz}$) at around $k_{x} = 0.098\times 10^{4}~\mathrm{cm}^{-1}$. However, the coupling is noticeably weaker than in the bulk case, as evidenced by the reduced anticrossing gap. This is primarily due to the surface DPPP being further detuned from the AFM magnon resonance, diminishing the magnonic contribution to the hybridized mode and thus lowering the overall interaction strength.

In contrast, at a slightly higher Fermi energy [Fig.~\ref{figure-blueshift-redshift-alpha-all-cases-separadas}\subref{figure-blueshift-redshift-alpha-B}], the surface polariton intersects the magnon resonance at a smaller in-plane wavevector, $k_{x} = 0.082\times 10^{4}~\mathrm{cm}^{-1}$. The resulting anticrossing is substantially larger, indicating stronger hybridization. This enhancement arises because the surface DPPP is now nearly resonant with the magnon mode in FeF$_2$, located around $\omega = 1.62~\mathrm{THz}$ and $k_{x} = 0.080\times 10^{4}~\mathrm{cm}^{-1}$, as highlighted by the intersection with the steep yellow dispersion curve in Fig.~\ref{figure-blueshift-redshift-alpha-all-cases-separadas}\subref{figure-blueshift-redshift-alpha-B}.

We now turn our attention to the influence of the bi-isotropic coefficient $\alpha$ on the dispersion of the Dirac plasmon--phonon polariton (DPPP), considering the case where the Fermi energy is fixed at $E_{F} = 1.5~\mathrm{eV}$. Interestingly, we observe that the inclusion of a finite $\alpha$ leads to a redshift in the DPPP dispersion, effectively bringing it closer to the dispersion observed in the uncoupled case with a lower Fermi energy of $E_{F} = 1.2~\mathrm{eV}$. This behavior suggests that the bi-isotropic parameter not only modifies the electromagnetic modes' interactions but also indirectly influences the effective carrier response involved in surface plasmon excitation.

From a physical perspective, the non-null bi-isotropic $\alpha$ parameter alters the constitutive relations by introducing a magnetoelectric-like cross-coupling between the electric and magnetic fields. This coupling modifies the effective impedance of the TI interface and can influence both the field confinement and the modal propagation constants. As a consequence, the electromagnetic mode associated with the Dirac surface states becomes less efficient in coupling to the charge carriers, effectively reducing the participation of high-momentum electrons in the surface mode. This manifests as a reduced plasmonic response, similar to what one would expect from a system with a lower carrier density, or equivalently, a lower Fermi level.

In other words, while a higher Fermi energy typically results in a blueshift of the DPPP due to increased free carrier density, the presence of a nonzero $\alpha$ appears to counteract this effect. The physical mechanism can be attributed to the redistribution of electromagnetic energy at the interface, driven by the magnetoelectric response, which shifts the modal dispersion toward lower frequencies. This redshift implies that the hybridized mode has a diminished electric field overlap with the high-density surface states, thereby mimicking the effect of a reduced Fermi level. Therefore, the bi-isotropic parameter $\alpha$ emerges as an additional degree of control over the light--matter interaction in TI/AFM heterostructures. It offers a novel route for tuning the dispersion properties of surface modes, enabling dynamic control over their spectral position and coupling strength. This tunability could be particularly useful for the design of THz devices where selective enhancement or suppression of specific polaritonic modes is desired.

\section{\label{section-final-remarks}Final Remarks}

In this work, we have investigated by effects of bi-isotropic parameters in the formation of hybrid surface polaritons in bilayer configurations. To this end, we have considered a bilayer constituted with TI medium endowed with bi-isotropic constitutive relations and an AFM medium. 

Using a scattering matrix formalism, we derived general dispersion relations that explicitly include $\alpha$, enabling us to analyze its influence on the hybridization between surface plasmons, optical phonons, and magnons. Our results show that increasing $\alpha$ leads to a pronounced redshift of the upper polaritonic branch and suppresses the characteristic anticrossing features, indicating a weakening of the hybrid interaction, possibly due to saturation or detuning effects. We also quantified the dependence of the coupling strength on the Fermi energy, and consequently on the carrier concentration, of the Dirac plasmon at the surface of the TI and at the TI/AFM interface. As the Fermi level increases, all polaritonic branches experience a blueshift due to the enhanced plasmonic response from the higher density of surface carriers, bringing the plasmon mode into closer resonance with the AFM magnon and strengthening the hybridization. Interestingly, the redshift induced by a finite $\alpha$ can partially offset this blueshift, effectively restoring the system to a weak-coupling regime. These findings reveal that the bi-isotropic parameter and the Fermi energy provide independent and complementary control over the spectral position and coupling strength of DPPMPs. Beyond offering insights into topologically mediated light--matter interactions, our results may open new possibilities for engineering tunable polaritonic phenomena in the terahertz regime.

Future investigations could generalize the present approach by replacing the scalar bi-isotropic parameter $\alpha$ with a full magnetoelectric tensor $\alpha_{ij}$, thereby capturing the anisotropic and direction-dependent nature of magnetoelectric interactions in realistic materials. This extension would lead to more complex constitutive relations and give rise to richer dispersion behaviors of DPPMPs in heterostructures composed of topological insulators and anisotropic magnetoelectric media. The resulting framework would enable the derivation of more general dispersion relations, potentially revealing new classes of hybrid excitations and offering a possible route toward understanding symmetry-governed light–matter interactions at terahertz frequencies.

\section*{Acknowledgments}  The authors thank FAPEMA, CNPq, and CAPES (Brazilian research agencies) for their invaluable financial support. P.D.S.S. is grateful to FAPEMA APP-12151/22. Furthermore, we are indebted to CAPES/Finance Code 001 and FAPEMA/POS-GRAD 04755/24.

\end{document}